\documentstyle[12pt]{article}

\catcode`\@=11
\def\marginnote#1{}
\newcount\hour
\newcount\minute
\newtoks\amorpm
\hour=\time\divide\hour by60
\minute=\time{\multiply\hour by60 \global\advance\minute
by-\hour}\edef\standardtime{{\ifnum\hour<12
\global\amorpm={am}%
        \else\global\amorpm={pm}\advance\hour by-12 \fi
        \ifnum\hour=0 \hour=12 \fi
        \number\hour:\ifnum\minute<10
0\fi\number\minute\the\amorpm}}
\edef\militarytime{\number\hour:\ifnum\minute<10
0\fi\number\minute}

\def\draftlabel#1{{\@bsphack\if@filesw {\let\thepage\relax
   \xdef\@gtempa{\write\@auxout{\string
      \newlabel{#1}{{\@currentlabel}{\thepage}}}}}\@gtempa
   \if@nobreak \ifvmode\nobreak\fi\fi\fi\@esphack}
        \gdef\@eqnlabel{#1}}
\def\@eqnlabel{}
\def\@vacuum{}
\def\draftmarginnote#1{\marginpar{\raggedright\scriptsize\tt#1}}
\def\draft{\oddsidemargin -.5truein
        \def\@oddfoot{\sl preliminary draft \hfil
        \rm\thepage\hfil\sl\today\quad\militarytime}
        \let\@evenfoot\@oddfoot \overfullrule 3pt
        \let\label=\draftlabel
        \let\marginnote=\draftmarginnote

\def\@eqnnum{(\theequation)\rlap{\kern\marginparsep\tt\@eqnlabel}%
\global\let\@eqnlabel\@vacuum}  }

\def\numberbysection{\@addtoreset{equation}{section}
        \def\theequation{\thesection.\arabic{equation}}}

\def\underline#1{\relax\ifmmode\@@underline#1\else
 $\@@underline{\hbox{#1}}$\relax\fi}

\catcode`@=12
\relax

\numberbysection
\newtheorem{definition}{\bf Definition}
\newtheorem{theorem}{\bf Theorem}
\newtheorem{corollary}{\bf Corollary}

\newtheorem{conjecture}{\bf Conjecture}
\def\qed{{\bf Q.E.D. }}
\def\proof{\noindent {\sl Proof:  }}
\def\Wr{\hbox{Wr}\>}
\def\fin{\end{document}}
\def\beq{\begin{equation}}
\def\eeq{\end{equation}}
\def\bea{\begin{eqnarray}}
\def\eea{\end{eqnarray}}
\def\nnn{\nonumber \\}

\def\Ab{{\bar A}}
\def\Bb{{\bar B}}
\def\Fb{{\bar F}}
\def\rhob{{\bar \rho}}
\def\Xb{{\bar X}}
\def\xb{{\bar x}}
\def\fb{\bar f}
\def\partialb{\bar \partial}
\def\Omegab{{\bar \Omega}}
\def\p{\Phi}
\def\xib{\bar \xi}
\def\kappab{\bar \kappa}
\def\Mb{\bar M}
\def\Xs{{\check X }}
\def\fs{{\check    f}}
\def\Fs{{\check  F}}

\def\Xsb{{ \skew4\check {\bar X  }}}
\def\sb{\bar s}

\def\wtE{{\widetilde E}}
\def\wtH{{\widetilde H}}
\def\wth{{\widetilde h}}
\def\Gcc#1 {{\cal G}^{[ #1 ]}_\|}
\def\Gcp#1 {{\cal G}^{[ #1 ]}_\bot}
\def\Cco#1 {{\cal C}^{[ #1 ]} }
\def\zb{\bar z}
\begin{document}

\begin{titlepage}
\nopagebreak
\begin{flushright}
LPTENS-93/47\\
hepth@xxx/9312040
\\
LPTENS--93/47
\\
December 1993
\end{flushright}

\vglue 2.5  true cm
\begin{center}
{\large\bf
$W$--GEOMETRY OF THE  TODA SYSTEMS\\
ASSOCIATED WITH \\
 NON-EXCEPTIONAL SIMPLE LIE ALGEBRAS
}

\vglue 1  true cm
{\bf Jean-Loup~GERVAIS} and {\bf Mikhail V.  SAVELIEV}\footnote{
On leave of absence from the Institute for High Energy Physics,
142284, Protvino, Moscow region, Russia.
}\\
{\footnotesize Laboratoire de Physique Th\'eorique de
l'\'Ecole Normale Sup\'erieure\footnote{Unit\'e Propre du
Centre National de la Recherche Scientifique,
associ\'ee \`a l'\'Ecole Normale Sup\'erieure et \`a l'Universit\'e
de Paris-Sud.},\\
24 rue Lhomond, 75231 Paris C\'EDEX 05, ~France.}

\medskip
\end{center}

\vfill
\begin{abstract}
\baselineskip .4 true cm
\noindent
The present paper describes  the $W$--geometry of the Abelian finite
non-period\-ic (conformal)  Toda
systems associated with the
$B,C$ and $D$ series of the simple Lie algebras endowed
with the canonical gradation. The principal tool here is a
generalization of the classical Pl\"ucker embedding of the $A$-case to the
flag manifolds associated with the fundamental representations of
$B_n$, $C_n$ and $D_n$, and a direct proof that the corresponding
K\"ahler potentials satisfy the system of two--dimensional finite non-periodic
(conformal)
Toda equations. It is shown that the $W$--geometry of the type
mentioned above
 coincide with the differential geometry of special
holomorphic (W) surfaces in target spaces which are  submanifolds (quadrics)
 of $CP^N$ with  appropriate choices  of $N$.
In addition,  these W-surfaces
are defined  to
satisfy quadratic holomorphic differential conditions that ensure consistency
of the generalized Pl\"ucker embedding. These conditions   are
automatically fulfiled  when  Toda equations hold.
\end{abstract}
\vfill
\end{titlepage}
\baselineskip .5 true cm
\section{Introduction}
A notion of  $W_{\cal G}$-geometry of ${\bf CP}^N$-target manifolds
associated with integrable systems,
recently invented in \cite{GM} for the case of $A_n$-Abelian Toda
system (see also \cite{G}) seems to be a very important tool for solvable
field theories as geometrical structures behind $W$-algebras,
as well as for algebraic and differential geometries themselves. In particular,
such a geometrical picture should be rather essential in the gauge
fields formulation of various models of the two--dimensional gravity, as well
as their generalizations for higher dimensions.
On the same footing as $W$-algebras, being the algebras of the characteristic
integrals ---conserved currents---
 for the corresponding nonlinear systems, guarantee, under appropriate
conditions, the integrability
property for these systems and give their classification, a description of
their $W$-geometry is equivalent, in a sense, to a classification scheme of
the corresponding  K\"ahler manifolds.\footnote{Some classification  of the
manifolds associated, in terms of the corresponding embedding problem
(Gauss--Codazzi and Ricci equations), with the Toda systems , and an attempt
of the geometrical formulation of the integrability criteria has been given
in \cite{S1}. However, it provides only some general links,
in particular, with
the diagonalisability of the corresponding $3$rd fundamental form, and is too
complicated for concrete conclusions.} It was shown in \cite{GM} that the
K\"ahler potentials of the intrinsic metrics induced on the corresponding
$W$-surfaces coincide with the $A_n$-Toda fields. In what follows we prove
that this fact takes place also for the $W$-surfaces associated  with all
other classical (non-exceptional) simple Lie algebras ${\cal G}$ and the
corresponding ${\cal G}$-Toda fields, and can be realized explicitly.
(In fact, we have conjectured this statement  already more than a year ago,
but only
now have a proof for that.) We believe that this notion is relevant for a wide
class of integrable dynamical systems as a geometrical counterpart of
$W$-algebras.

For the  readers who are not familiarized with the notions
$W$-algebra and $W$-geometry, at least in the meaning which will be used in
our paper, let us recall it in a few words.

By $A_n$--$W$-geometry we mean
the geometry of the   ${\bf CP}^n$ $W$-surface of ref.\cite{GM}  which
are   two--dimensional manifolds   ${\bf \Sigma}$ supplied with a complex
structure, and  an embedding into
${\bf CP}^n$ such that half of the coordinates $X^A,\,1\leq A\leq n+1$, of the
enveloping space holomorphically depends on a local coordinate $z$ of
${\bf \Sigma}$, $X^A=f^A(z)$, while the other half, $\bar{X}^A$, are
anti-holomorphic functions $\bar{X}^A=\bar{f}^A(\bar{z})$. In other words,
in the language of algebraic geometry, we speak here, with account of the
appropriate reality condition, about holomorphic
curves in the corresponding
projective target space ${\bf CP}^n$.  Note that
we  call them   surface, instead,
on account of their
 real dimension. This is more appropriate for
 applications to conformal models and
string theories.
 The corresponding $W$-surface is called
$W_{A_n}$-surface; it is related to the first  fundamental representation
of $A_n$, and there are associated surfaces related to the other fundamental
representations of $A_n$. As shown in ref.\cite{GM}, and
as we shall recall below,  the
$W_{A_n}$- and associated surfaces
correspond to the classical extrinsic geometry of the curves in ${\bf CP}^n$
having deal with the Pl\"ucker image of the Grassmannians ${\cal G}r(n+1|\, k)$
in ${\bf CP}^{{ {n+1}\choose k } -1}$.

 Since the corresponding complex projective target space is defined
as the quotient of the space ${\bf C}^{n+1}$ by the equivalence lifting
(local rescaling of the coordinates),  it immediately follows
from the given definition for the $W_{A_n}$-surface,
that these holomorphic (anti-holomorphic)
functions are  solutions of some homogeneous ordinary
differential equation of the $(n+1)$th order,
\begin{equation}
\frac{\partial ^{n+1}f^A}{\partial z^{n+1}}=\sum _{\alpha =1}^n W_{\alpha}
\frac{\partial ^{\alpha}f^A}{\partial z^{\alpha}},\quad
\frac{\partial ^{n+1}\bar{f}^A}{\partial \bar{z}^{n+1}}=
\sum _{\alpha =1}^n \bar{W}_{\alpha}
\frac{\partial ^{\alpha}\bar{f}^A}{\partial \bar{z}^{\alpha}}
\label{A0}
\end{equation}
with nonzero\footnote{In this paper, we do not consider
singular points of W-surfaces. This may be done straightforwardly.}
 coefficients $W_{\alpha}$ ($\bar{W}_{\alpha}$). In writing these
equations we made use of the fact that the Wronskians $\Wr[f(z)]$ and
$\Wr[\bar{f}(\bar{z})]$ constructed with the functions $f^A(z)$ and
$\bar{f}(\bar{z})$ do not vanish at regular points of $\Sigma$, and one can
divide by them. In other words, here we deal with the osculating hyperplanes
to the generic $W$--surfaces. Note also in this context that the
linear system of the Pl\"ucker quadrics which provides the decomposability
property of the Pl\"ucker image, is automatically satisfied on the class of
the solutions to Eqs.(\ref{A0}).

One of the main point of this paper is to
define  some submanifolds in ${\bf CP}^N$, with appropriate $N$,
 which are target  spaces
for the  $W_{\cal G}$-surfaces relevant for the other
complex simple Lie algebras ${\cal G}$. These spaces are
specified by quadratic conditions whose origin is as follows.
Now  the set of the indices $\alpha$
in the sum in the r.h.s. of (\ref{A0}), for which the coefficients
$W_{\alpha}\not= 0$ and $\bar{W}_{\alpha}\not= 0$, coincides with the values
of the exponents of the algebra ${\cal G}$. The vanishing of the
corresponding coefficient functions in the series in the r.h.s. in (\ref{A0})
leads to the set of the quadratic relations on the embedding functions $f^A(z)$
($\bar{f}(\bar{z})$) and their derivatives up to the $(n-1)$th order; and these
local conditions on the functions ensures
consistency of the
generalized Pl\"ucker embedding described  in the present paper.
It happens that the corresponding K\"ahler manifolds in $W$-geometry are
ultimately related to the Toda fields, being described by the equations of
the two--dimensional finite nonperiodic Toda system which, fortunately, are
exactly solvable
\cite{LS}.

In accordance with \cite{BG}, see also \cite{O'R},  the results of the Toda
theory provide a realization of the $W$-algebras in terms of the polynomials
constructed with the corresponding Toda fields, more exactly via their
derivatives. By $W$-algebra we mean\cite{BG}
 an algebra with the defining relations
\[
\{ W_{\alpha}(z), W_{\beta}(z')\}=\sum _a {\cal P}^a_{\alpha , \beta }
(W)\delta ^{(a)}(z_1-z_1') \]
just for the coefficient functions entering (\ref{A0}), which realize the
corresponding infinitesimal $W$-transformation of the functions $f^A$ and
$\bar{f}^A$.
Here ${\cal P}^a_{\alpha , \beta }$ are polynomials of the $W_{\alpha}$'s
and their derivatives over spatial variable $z_1$ in two--dimensional
space--time $\{ \bar{z}, z\}=\{ (z_0\pm z_1)/2\}$ with the
metric\footnote{One may also define $z$ as a complex variable. Then
$g_{11}=g_{22}=1$.}
$g_{11}=-g_{22}=1$; the Poisson brackets are taken for equal time value,
$z_0=z_0'$. Moreover, such objects as the elements $W_{\alpha}$ with values
in the ring of gauge invariant differential polynomials, are quite known in
the integrable systems business, being in fact local characteristic integrals
for the corresponding system of the partial differential
equations;\footnote{As for the characteristic integrals, a general method of
their explicit construction for a wide class of two--dimensional nonlinear
equations, in particular for the Abelian Toda system and its nonabelian
versions, is given in \cite{LS}.} the
existence of these integrals provides integrability of the system. So, the
theory of integrable systems is a natural place where Lie group--algebraic
and differential and algebraic geometry aspects are intersected as the
$W$-algebra\&$W$-geometry.

The relevant instrument for our description is a modification of the Pl\"ucker
embedding for all classical
series of the simple Lie algebras. Recall that the standard  Pl\"ucker
embedding is
formulated for the $A_n$-case, see e.g. \cite{GH}, and result in the
infinitesimal and global Pl\"ucker formulas.
Note that the statement  which
generalizes the infinitesimal Pl\"ucker formula (related to the canonical
distribution) for an arbitrary simple Lie algebra ${\cal G}$, has been
conjectured in \cite{Giv} and then proved in \cite{Pos}, see also \cite{Yo},
 without any connection with integrable
systems, $W$--geometry, and all that. In \cite{RS}, using the relevant
differential geometry setting, while without any reference to a coordinate
 representation of the corresponding flag manifolds associated with the Abelian
Toda system, there was also obtained  the generalized infinitesimal Pl\"ucker
formula for an arbitrary simple Lie algebra ${\cal G}$. These flag manifolds
are the quotient spaces $G/P$ with $P$ being the maximal nonsemisimple
parabolic subgroups of $G$, ${\cal G}=\mbox{ Lie }G$.

As we will show, the relevant $W$--manifolds in our approach are
related to the target manifolds of integrable systems gauged by a semi--direct
product of a nilpotent and semisimple subgroups of Lie ${\cal G}$. In other
words, the K\"ahler manifolds in question arise as parabolic spaces for
a simple Lie group $G$ whose coordinates satisfy some homogeneous equations
quadratic in the coordinates. Similar to those in the $A_n$-case, these
coordinates are some minors of a matrix representative of the corresponding
cosets; and they are submitted to  homogeneous quadratic
equations. However, only a part of these equations are quadrics of exactly
the Pl\"ucker type;\footnote{Recall that just the system of the Pl\"ucker
quadrics provides the condition of decomposability of a multivector in the
corresponding complex projective space, and hence defines the Pl\"ucker image
in it \cite{GH}.}
while the other ones are caused by the specific features of such
algebras as the orthogonal algebras. In general, the Toda fields are
related with (minors of) determinants. Thus a natural tool is
to perform skew products of representations;   hence we
shall use fermionic operators, see below. For $A_n$, one can obtain
all  finite dimensional irreducible representation by skew
products of a  finite number of copies of
the first fundamental one. As we will see, this is related with the fact
that, since the Dynkin diagram is a simple line, the derivation of
Toda equation goes rather smoothly from the first root  to the last.
For other  algebras, the situation is more complicated.
Now, skew products of the first fundamental representation are not
enough. One should also include the last one for $B_n$, and the last two
for $D_n$.
These representations are of a different nature,
and their highest weights have half
integer components. In the Dynkin diagram they correspond to
non-generic points with branching where the derivation of Toda
equations is much more subtle. All these difficulties  will be overcome  in
the ensuing discussion.
The study of the problem for the orthogonal algebras already contains
\footnote{The case of $C_n$ is much simpler.}
seemingly all peculiarities and ``underwater stones'' that are
naturally absent
in the case of $A_n$. Thus we believe that our picture is truly general.

\label{note}
To clarify the principal difference in the formulation of the problem in
question for the simple Lie algebras other than $A_n$, we
shall first recall  some results for the $A_n$-case,
 mainly following\footnote{Part of this discussion already appeared
in the preprint version
 of the second article of  ref.\cite{GM}. It was removed from
the printed version in order to shorten the article and  satisfy   the
editor's request.} ref.\cite{GM};
and complete them by some reasonings leading to the Pl\"ucker quadratic
relations. As already emphasized,
and in distinction to the case of the Lie algebra $A_n$, a similar study of
$W_{\cal G}$-geometry of the Toda systems for other simple Lie algebras
is not so direct if one wishes  to realize the program in the coordinate basis
explicitly.

Note that   an important instrument of our consideration
will be   fermionic realizations
of the elements of the classical Lie algebras, similar to the one
which has been efficiently used in \cite{GM} for an investigation of the
$W$-geometry of the $A_n$-Abelian Toda systems. The  main advantage of this
realization, apart from its technical simplicity, is that it allows to
interpolate between different fundamental representations, and relate their
basis vectors. This is extremely suitable for a solution of the problem under
consideration, where skew products of representations are the
key.

We give an explicit formulation of a relevant modification
of the Pl\"ucker mapping for the manifolds
associated with the fundamental representations of an arbitrary classical Lie
algebra ${\cal G}$, and a direct proof that the corresponding K\"ahler
potentials satisfy the system of partial differential equations of the Toda
type. In general, we believe\footnote{This was also hypothesized in
 the first article of
ref.\cite{GM}.} that every integrable system is naturally
associated with the corresponding K\"ahler manifold ---that means with the
relevant group $G$ and its gauging--- which in turn is determined by the
invariance subgroup for the choosen representation space. Here the manifold is
defined by the gradation of the Lie algebra ${\cal G}$ and the grading spectrum
of the corresponding component of the Maurer-Cartan 1-form which results in
the nonlinear systems in question. So, the algebraic counterparts of the given
$W$--manifold are Lie algebra, its gradation, and the grading spectrum of the
connections. Note also that, as we have understood from  discussions with
M. Kontsevich and Yu. I. Manin, our consideration of the nonlinear Toda type
systems as holomorphic curves in the corresponding projective spaces, seems
to be closely related to variations of the Hodge structures in the spirit of
Ph. A. Griffiths.

\bigskip

\section{Abelian Toda systems, and $W$--$A_n$-geometry}
Consider a finite--dimensional complex simple Lie algebra ${\cal G}\equiv
{\cal G}$ of rank $n$, with the following defining relations
\begin{equation}
{}[h_i, h_j]=0,\;[h_i, E_{\pm j}]=\pm K^{\cal G}_{ji}E_{\pm j},\; [E_{+i},
E_{-j}]=\delta _{ij}h_i\label{A1}
\end{equation}
for its Cartan $\{ h_i\}$ and Chevalley $\{ E_{\pm i}\}$ elements, $1\leq i
\leq n$; and $K^{\cal G}$ being the Cartan matrix of
${\cal G}$. Let ${\cal G}$ be
endowed with the canonical gradation,
\[
{\cal G}=\bigoplus_{m\in {\bf Z}}{\cal G}_m,\;[{\cal G}_m, {\cal G}_r]
\subset {\cal G}_{m+r},\]
for which ${\cal G}_0=\{ h_i\}$ is Abelian, and ${\cal G}_{\pm 1}=\{ E_{\pm
i}\}$. Then, in accordance with the group--algebraic approach
\cite{LS},\footnote{Here and in what follows, the results concerning the
Toda theory are given following \cite{LS}.} the
zero--curvature condition
\[
{}[\partial /\partial \bar{z}+A_+, \partial /\partial z+A_-]=0 \]
for the connection components $A_{\pm}(z,\bar{z})$  taking values
in the subspaces ${\cal G}_0\bigoplus {\cal G}_{\pm 1}$, respectively, results
in the partial differential equations describing two--dimensional finite
nonperiodic Toda system
\begin{equation}
\partial \bar\partial \Phi _i=-\exp \rho_i, \;1\leq i \leq n
;\; \rho_i=\sum_{j=1}^nK^{\cal G}_{ij}\Phi _j; \label{A2}
\end{equation}
$\partial \equiv \partial/\partial z,\;\bar{\partial} \equiv
\partial/\partial \bar{z}$. The general solution to this system is written,
in one of the
equivalent forms, as follows. Associated with each fundamental
representation --  with highest weight
$\lambda_k$ --
 there exists  a Toda
field $\p_k$ defined by
\begin{equation}
e^{-\Phi_k}= <\lambda_k | \Mb^{-1}(\zb) M(z) |\lambda_k >
e^{-\xib_k(\zb) -\xi_k(z)}, \quad \hbox{where}
\label{A3}
\end{equation}
\begin{eqnarray}
{ d M \over dz}=M \sum_{j=1}^n s_j(z)E_{-j}&&, \quad
{d \Mb \over d\zb}=\Mb \sum_{j=1}^n \sb_j(z)E_{j}, \nonumber
\\
\xi_k(z)=\sum_{j=1}^n  ((K^{\cal G})^{-1})_{kj}
 \ln s_j(z)&&,\quad
\xib_k(\zb)=
\sum_{j=1}^n ((K^{\cal G})^{-1})_{kj} \ln \sb_j(\zb).\label{15.45}
\end{eqnarray}
The n  functions $s_j(z)$ and $\sb_j(\zb)$ which are arbitary
will
be called screening functions, since they are the classical
analogues of the Coulomb-gas operators.
They determine the
general solution of the Goursat (boundary) value problem for (\ref{A3}).
The matrix element $<\lambda _k|\bar{M}^{-1}M|\lambda _k>$ in the r.h.s. of
(\ref{A3}), is in fact, the tau--function for system (\ref{A2}), associated
with the
highest weight vector matrix element
of the $k$--th fundamental representation of ${\cal G}$ with the
highest state $|\lambda _k>$.

For later use, we note that one can consider the same formulae as above, but
for irreducible representations which are not fundamental. By definition
of the fundamental weights, any highest weight is of the form $\lambda =
\sum _k\nu _k\lambda _k$, where $\nu _k$ are nonnegative integers. Then the
generalization of Eqs.(\ref{A3}) and (\ref{15.45}) is
\begin{equation}
e^{-\sum _k\nu _k\Phi _k}=e^{-\sum _{j=1}^n(\vec \lambda \cdot \vec \lambda
_j)\mbox{ ln }s_j(z)-\sum _{j=1}^n(\vec \lambda \cdot \vec \lambda
_j)\mbox{ ln }\bar{s}_j(\bar{z})}<\lambda
|\bar{M}^{-1}(\bar{z})M(z)|\lambda >.
\label{A3b}
\end{equation}
The proof of this equality is carried out by taking the corresponding powers
of (\ref{A3}). This automatically constructs the highest weight vector with
highest weight $\lambda$, on which the action of $M$ and $\bar{M}$ can be
derived solely from the Lie group theory.

For the case of the $A_n$-Toda system, all the Toda fields
$\exp (-\Phi _j), \,
1\leq  j\leq  n$, are expressed via the first one,
$\exp (-\Phi _1)$ which can
be written as
\begin{equation}
\exp (-\Phi _1)=\sum _A f^A(z)\cdot \bar{f}^A(\bar{z}).
\label{A5}
\end{equation}
Here the functions $f^A(z)$ and
$\bar {f}^{\bar {A}}(\bar{z})$  satisfy
the conditions which can be expressed in terms of
the Wronskians constructed with these functions,
\begin{equation}
\Wr[f(z)]=1,\; \Wr[\bar {f}(\bar{z})]=1;
\label{A6}
\end{equation}
and are formulated via the independent (chiral) screening functions
$s_i(z)$
and $\bar{s}_i(\bar{z})$,  $1\leq i\leq n$, entering the general solution
(\ref{A3})  as the nested integrals (\ref{15.50}), in our, $A_n$-case with
$i_s=s$.
All other Toda fields  $\exp (-\Phi _j), \, j \, >\, 1,$ are written
in terms of $\exp (-\Phi _1)$ by the formulas
\begin{equation}
\exp (-\Phi_j)= \Delta _j,
\label{A9}
\end{equation}
and hence
\begin{equation}
\exp (-\Phi_j)=\sum _j \mbox {det} _j(f) \cdot \mbox {det} _j(\bar f).
\label{A10}
\end{equation}
Here sum runs over all the j-th order minors $\mbox {det} _j(f)$ and
$\mbox {det} _j(\bar f)$ constituted by the first $j$ rows of the matrices
$(\partial )^Bf^A$ and $(\bar {\partial })^B\bar {f}^A$, respectively;
$\Delta _j$ is the j-th order principal minor of the matrix
$\partial ^A\bar {\partial }^B \exp (-\Phi _1)$. Recall that such minors
satisfies very important relation
\begin{equation}
\partial \bar{\partial }\mbox{ log }\Delta _j=
\frac{\Delta _{j+1}\cdot \Delta _{j-1}}{\Delta _j^2},
\label{A11}
\end{equation}
which is used in what follows.

To have a more precise picture of what we are going to do in the general case,
let us reproduce here some basic steps leading to the $W_{A_n}$-geometry.
We mainly follow (see footnote on page \pageref{note})
the paper \cite{GM}, but  supply
some additional formulas needed for understanding the quadratic relations
of the Pl\"ucker type which are absent there. First introduce
the relevant notations, see e.g. \cite{B}, \cite{Gil}. Let $\vec e_p$ be the
a set of orthonormal vectors in a $n+1$ Euclidean space
 $\vec e_p\cdot \vec e_q=\delta_{pq},\, 1\leq p,q\leq
n+1$, which parametrize the positive and negative roots
$\pm (\vec e_p-\vec e_q),\; 1\leq p<q\leq
n+1$, of $A_n$The vectors
 $\vec \pi _i=\vec e_i-\vec e_{i+1}$ are a set of simple
roots. Denote by
$\vec \lambda _i, \,1\leq i\leq n$, the fundamental weights of $A_n$,
$\vec \lambda _i$  is equal to
$\sum _{j=1}^i\vec e_j-i\sum _{j=1}^{n+1}\vec e_j/(n+1)$. The
corresponding highest
weight state
$|\lambda _i>$  is defined by the
conditions
\begin{equation}
h_j|\lambda _i>=\delta _{ij}|\lambda _i>, \qquad E_{+j}|\lambda _i>=0 \;
\mbox{ for all } 1\leq j\leq n; \label{A14}
\end{equation}
and normalization $<\lambda _i|\lambda _i>=1$; moreover, $E_{-j}|\lambda _i>
=0$ for all $j\not= i$. In accordance with this definition, the whole
representation space  of the $i$-th fundamental representation consists of all
the vectors $|A_p\cdots A_1>_i\; \equiv E_{-A_p}\cdots E_{-A_1}|\lambda _i>$,
$1\leq A\leq n,\, 0\leq p\leq N_i-1,\, N_i={{n+1} \choose i }$, with
nonzero norm. In what follows we use the fermionic realization of the elements
of $A_n$, in which the Cartan and Chevalley generators are written as
\begin{equation}
h_i=\flat_i^+\flat_i-\flat_{i+1}^+\flat_{i+1},\; E_i=\flat_{i+1}^+\flat_i,
\; E_{-i}=\flat_i^+\flat_{i+1},\; 1\leq i\leq n+1.\label{A15}
\end{equation}
Here $\flat_p$ and $\flat_p^+$ are fermionic operators satisfying the
standard anticommutation relations
\[
{} [\flat_p, \flat_q]_+= [\flat_p^+, \flat_q^+]_+=0,\; [\flat_p,
\flat_q^+]_+=\delta_{pq};\]
and there exists a vacuum state $|0>$, such that $\flat_p|0>=0$ for all
$1\leq p\leq n+1$, and, correspondingly, for the dual state $<0|\flat_p^+=0$.
In this, the $i$-th particle state, which is the highest weight state of the
$i$-th fundamental representation of $A_n$, is obtained from the vacuum
(cyclic)
vector by the action of the raising operators, namely
\[
|\lambda _i>=\flat_i^+\flat_{i-1}^+\cdots \flat_1^+|0>, \mbox{ and,
respectively,} <\lambda _i|=<0|\flat_1\cdots \flat_{i-1}\flat_{i}.\]

Now, let us consider the coset space ${\cal C}^{[i]}$ associated with the
$i$-th fundamental representation of $A_n$. It is quite clear that here the
vector space of $A_n$ is splitted into the semi--direct sum ${\cal G}=
{\cal G}_{\|}^{[i]}\uplus {\cal G}_{\bot}^{[i]}$
of the subalgebra ${\cal G}_{\|}^{[i]}$
which is the stabiliser of the highest weight state $|\lambda _i>$,
\begin{equation}
{\cal G}_{\|}^{[i]}  =\{ h_j,\, j\not= i; \;
E_{-\vec e_p+\vec e_q},\, q\leq p\leq i
\mbox{ and } q\geq i+1;\;
E_{\vec e_p-\vec e_q},\, p<q\} ;
\label{A16}
\end{equation}
and the complement
\begin{equation}
{\cal G}_{\bot}^{[i]}  =\{ h_i; \; E_{-\vec e_p+\vec e_q},\, q\leq i,
\mbox{ and } p\geq i+1 \}. \label{A17}
\end{equation}
 In these notations
the coset ${\cal C}^{[i]}$ is constructed by exponentiation of the linear span
of ${\cal G}_{\bot}^{[i]}$, namely
\begin{equation}
{\cal C}^{[i]}\, : \, e^{\Omega _i}|\lambda _i>=
\sum _{A_1< A_2< \cdots < A_i}
\Lambda ^{[i]}_{A_1,\cdots , A_i} \flat_{A_i}^+ \cdots \flat_{A_1}^+|0>,
\label{A18}
\end{equation}
where
\begin{equation}
\Omega _i=\kappa _i(\flat_i^+\flat_i-\flat_{i+1}^+\flat_{i+1})+
\sum_{1\leq q\leq i,\; i+1\leq p\leq n+1}x^{[i]}_{pq}\flat_p^+\flat_q.
\label{A19}
\end{equation}
Every finite dimensional irreducible representation
of $A_r$ (and hence all the fundamental ones) is contained in a skew  product
of the finite number of copies of the $1$st fundamental irreducible
representation. This is why we may obtain all fundamental representations
in this way. In practice, the calculation of the coordinates
$\Lambda ^{[i]}_{A_1,\cdots , A_i}$ goes in two steps. First one may introduce
quantities noted $X^{[p], \alpha \, A_p}$, with $1\leq p \leq n$ defined by
\begin{equation}
e^{\Omega _i}\flat_\alpha^+e^{-\Omega _i}=
\sum _{A=1}^{n+1}X^{[i], \alpha,  A}
\flat_A^+ \quad \mbox{ for } A\leq i;
\label{A20p}
\end{equation}
and Eq.\ref{A18} gives
\begin{equation}
\sum _{B_1< B_2< \cdots < B_i}
\Lambda ^{[i]}_{B_1,\cdots , B_i} \flat_{B_i}^+ \cdots \flat_{B_1}^+|0>=
\sum _{A_1,\,     \cdots\,   A_i}
X^{[i]i,\, A_i}
\cdots X^{[i]1,A_1}\flat^+_{A_i}\cdots \flat^+_{A_1}|0>,
\label{A19p}
\end{equation}
Thus the coordinates $\Lambda ^{[i]}_{A_1,\cdots , A_i}$ of the
 $i$-th fundamental-representation space are the
antisymmetrized combinations of the products $X^{[i]i,A_i}X^{[i]i-1,A_{i-1}}
\cdots X^{[i]1,A_1}$.
However, for the sake of brevity, we conventionally
call the quantities $X^{[i]\alpha ,A}$  coordinates also.

The explicit relation  the coordinates $\xi _i,\, x^{[i]}_{pq}$ of the
coset ${\cal C}^{[i]}$ and the coordinates $X^{[i]\alpha ,A}$ defined on the
$i$-th fundamental representation space, is obtained by the straightward
computation using the simple
formula
\begin{equation}
e^{\Omega _i}\flat_A^+e^{-\Omega _i}=\sum _{B=A}^{n+1}u^{[i]}_{Bi}
\flat_B^+ \quad \mbox{ for } A\leq i; \label{A20}
\end{equation}
where
\beq
u^{[i]}_{i\, i}=e^{\kappa _i},\quad u^{[i]}_{i+1\, i}=\frac{\sinh
\kappa _i}{\kappa _i}x^{[i]}_{i+1,i},\quad u^{[i]}_{p\, i}=
\frac{e^{\kappa _i}-1}{\kappa _i}
x^{[i]}_{p,i},\quad i+2\leq p\leq n+1;
\eeq
and for  $A<i$, $A<B\leq i$, and $i+2\leq p\leq n+1$,
\beq
u^{[i]}_{AA}=1,\quad u^{[i]}_{BA}=0,\quad
u^{[i]}_{i+1\, A}=\frac{1-e^{-\kappa _i}}{\kappa _i}x^{[i]}_{i+1,A},\quad
u^{[i]}_{p\, A}=x^{[i]}_{p,A}.
\eeq
Then we come to the following parametrization
of the coordinates $X^{[i]\alpha , A}$ entering
(\ref{A18}),
\begin{equation}
 X^{[i]i, A}=u^{[i]}_{Ai} \quad \mbox{ for }A\geq i,\quad
 X^{[i]i, A}=0 \quad \mbox{ for }A\leq i-1;\label{A23}
\end{equation}
and for $\alpha <i$,
\begin{equation}
 X^{[i]\alpha , A}=u^{[i]}_{A\alpha} \quad \mbox{ for }
\; A\geq i+1,\;
X^{[i]\alpha , \alpha}=1, \label{A24}
\end{equation}
\[
 X^{[i]\alpha , A}=0 \quad \mbox{ for }
 \; 1\leq A\leq i-1,\; A\not= \alpha.\]
With these coordinates, representation (\ref{A18}) leads to the corresponding
K\"ahler potential ${\cal K}^{[j]}$ of the manifold ${\cal C}^{[j]}$,
\begin{equation}
{\cal K}^{[j]}=\mbox{ log }\| \Lambda ^{[j]}\| ^2\equiv
\mbox{ log }[<0|{\bar{X}}^{[j]1}\cdots {\bar{X}}^{[j]j}
{X}^{[j]j}\cdots {X}^{[j]1}|0>].
\label{A25}
\end{equation}
Here
\[
{X}^{[j]\alpha}\equiv \sum _AX^{[j]\alpha ,A}\flat_A^+,\;
{\bar{X}}^{[j]\alpha}\equiv \sum _A\bar{X}^{[j]\alpha ,A}\flat_A;\;
\Lambda ^{[j]}\equiv \sum _{A_1<\cdots <A_j}\Lambda ^{[j]}_{A_1,\cdots
,A_j}\flat_{A_j}^+\cdots \flat_{A_1}^+|0>.\]
Note that, in fact, formula (\ref{A25}), with account of the aforementioned
identification
of the K\"ahler potentials ${\cal K}^{[j]}$ with the Toda fields $\Phi _j$,
which is discussed below,
 gives a different representation of the
tau--function for Eqs.(\ref{A2}) than those from \cite{LS}. Namely, it
is more adequate to the
skew--product structure of the fundamental representation space.

The relation Eq.\ref{A19p} between the
coordinates $\Lambda ^{[i]}_{A_1,\cdots , A_i}$ of the $i$-th fundamental
representation space, and the parameters $X^{[i] \alpha, A}$
immediately gives  the   quadratic relations
\begin{equation}
\sum _{\omega}\delta _{\omega}\,\Lambda ^{[i]}_{A_1,\cdots , A_{i-1}
A_{\omega (i)}}\,\Lambda ^{[i]}_{A_{\omega (i+1)},\cdots ,A_{\omega (2i)}}
=0,\quad 1\leq A_1<\cdots <A_{2i}\leq n+1,
\label{A26}
\end{equation}
that defines Pl\"ucker quadrics;
 and the same is for $\bar{\Lambda }^{[i]}_{A_1,
\cdots , A_i}$. Here the sum runs over all inequivalent permutations
$\omega$ of the integers $i,i+1,\cdots ,2i$, with account of the
antisymmetricity of $\Lambda ^{[i]}_{A_1,\cdots , A_i}$ under permutations
of the indices $A_1,\cdots , A_i$, and $\delta _{\omega}$ is the parity of
the permutation $\omega$. Note that the system of quadrics (\ref{A26}) comes,
in accordance with the contraction procedure given in \cite{GH}, from
equating to zero the skew--product of the $i$-vector $\hat{\Lambda}^{[i]}$,
$\Lambda ^{[i]}=\hat{\Lambda}^{[i]}|0>$, and the vector
${}^{-(i-1)}\hat{\Lambda}^{[i]}$ which in turn is obtained from
$\hat{\Lambda}^{[i]}$ under the action of $(i-1)$ annihilation operators
$\flat$.

Indeed, one can get convinced that the metric arised from such K\"ahler
potential (\ref{A25}), constructed with the coordinates $X^{[j]\alpha ,A}$
and $\bar{X}^{[j]\alpha ,A}$, is invariant under their transformation by an
arbitrary $j\times j$ matrix from $GL(j, {\bf C})$, as well as
relations (\ref{A26}). In particular, for the case of the 1st
fundamental representation of $A_n$, the corresponding complex projective
target space is defined as the quotient
of the space ${\bf C}^{n+1}$ by the equivalence rescaling
${\bf X} \sim {\bf Y}, \mbox{ if } X^A=Y^A\rho (Y), \mbox{ and }
\bar {X}^{\bar {A}}=\bar {Y}^{\bar {A}}\bar {\rho }(\bar {Y})$;
with arbitrary chiral functions $\rho (Y)$ and $\bar {\rho }(\bar {Y})$;
the metric, invariant under this rescaling, is the Fubini--Study metric
corresponding to the
K\"ahler potential ${\cal K}^{[1]}=\mbox{ ln }\sum _{A=1}^{n+1}
X^{A}\bar {X}^{A}$.

On the $W_{A_n}$-surface ${\cal K}^{[1]}=-\Phi _1$, and on the associated
surfaces the K\"ahler potentials also coincide, up to the sign, with the
corresponding $A_n$-Toda fields which are given by the r.h.s. of expressions
(\ref{A10}).  So, as we have already mentioned,
a relevant object for the description of the $W_{A_n}$-geometry of
${\bf C}^n$ - target manifolds with a positive curvature form is the Pl\"ucker
embeddings of the Grassmannians ${\cal G}r(n+1|\, k)$ in the projective spaces,
${\cal G}r(n+1|\, k)\Rightarrow {\bf CP}^{{ n+1 \choose k }\, -1}$. In this,
we identify $X^{[1]1,A}$ and $\bar{X}^{[1]1,A}$ with the
embedding functions in (\ref{A5}) with account of condition (\ref{A6}), and,
in general, putting
\begin{equation}
X^{[i]\alpha ,A}= \partial ^{\alpha -1}f^A, \qquad
\bar{X}^{[i]\alpha ,A}= \bar{\partial }^{\alpha -1}\bar{f}^A,\label{A27}
\end{equation}
leads to  the aforementioned equalities  between
the K\"ahler potentials and
the $A_n$-Toda fields\cite{GM} (up to a minus sign). Then, the known
Pl\"ucker representation for the pseudo--metrics $d{\cal S}^2_k$ on $\Sigma$,
specified by the corresponding K\"ahler potentials, see e.g. \cite{GH}, is
written in terms of these fields as
\begin{equation}
d{\cal S}^2_k\,=\, \frac{i}{2}\exp 2(-\Phi _{k-1}+2\Phi _k-\Phi _{k+1})\,
dz\, d\bar {z}\,\equiv \,\frac{i}{2}\exp 2(\sum
_{j=1}^nK^{A_n}_{kj}\Phi _j)\,dz\, d\bar {z},
\label{A31}
\end{equation}
where $K_{ij}^{A_n}$ is the Cartan matrix of the algebra $A_n$.

In distinction to the case of the Lie algebra $A_n$, a similar study of
$W_{\cal G}$-geometry of the Toda systems for other simple Lie algebras
requires a modification of the standard Pl\"ucker embedding. The point is
that  the number of independent screening functions $s_i(z)$ and
$\bar{s}_i(\bar{z})$  determining the general solution (\ref{A3}) of the
corresponding  Toda systems is always $2n$, while the number of
functions  $f^A(z)$ and $\bar{f}^{\bar{A}}(\bar{z})$ is, in general,
much more, even for
the representations of ${\cal G}$ of minimal dimension.
 For example, the relations
$\Phi_j=\Phi_{n-j+1}$ for the Toda fields of $A_n$ leads for $n=2s$ to the
solutions corresponding to the series $B_s$, for $n=(2s-1)$ -- to those of
$C_s$; and in terms of the screening functions this equating means that
$s_i(z)=s_{n-i+1}(z)$ and  $\bar{s}_i(\bar{z})=\bar{s}_{n-i+1}(\bar{z})$.
Moreover, due to the known relation between the orders of the functionally
independent characteristic integrals for the Toda systems, in other words
the $W$ - elements, and those of the Casimir operators of the corresponding
Lie algebras ${\cal G}$, the mentioned difference is also quite natural.
\section{The target spaces as  group-orbits of fundamental weights:
the case of  $D_n$.}
\label{Dn}
\subsection{Fermionic realizations}

Let us first recall some properties of the Lie algebra
$D_n$, see e.g. \cite{B}, \cite{Gil},  and its fermionic
realization.
The roots  are of the form
\beq
\vec \alpha  = \pm \vec e_i \pm \vec e_j,\>
\hbox{with}\> 1\leq i<j\leq n,
\label{15.1}
\eeq
in the $n$-dimensional space
span by the orthonormal vectors $\vec e_i$. The elements of $D_n$ can be
realized using  $2n$
fermionic operators $\flat_{\pm j}$, $j$ $=1$, $\cdots $ $n$, which
satisfy the relations
\beq
\bigl [ \flat_\ell, \flat_m\bigr ]_+=
\bigl [ \flat^+_\ell, \flat^+_m\bigr ]_+=0, \quad
\bigl [ \flat_\ell, \flat^+_m\bigr ]_+=\delta_{\ell, \, m},
\quad -n\leq \ell, \, m \leq  n;
\label{anticom}
\eeq
as
\beq
E_{\vec e_i\mp \vec e_j}=
\flat_i^+\flat_{\pm j} -\flat^+_{\mp j} \flat_{-i},
\quad
E_{-\vec e_i\pm \vec e_j} =E_{\vec e_i\mp e_j}^+
= \flat_{\pm j}^+ \flat_i - \flat_{-i}^+ \flat_{\mp j},
\label{15.2}
\eeq
and for the  Cartan generators
\beq
h_i=H_i-H_{i+1}, \quad i=1,\, \cdots ,\, n-1,
\quad h_n=H_{n-1}+H_n; \quad
H_i=\flat_i^+\flat_i- \flat_{-i}^+\flat_{-i}.
\label{15.5}
\eeq
A set of  simple
positive roots  is $\vec \pi_i=\vec e_i-\vec e_{i+1}$, $i=1$,
$\cdots $ $n-1$, and $\vec \pi_n=\vec e_{n-1}+\vec e_{n}$.
Let $E_{\pm i}$ be the corresponding Chevalley elements.
One has
\beq
E_i=\flat_i^+\flat_{i+1} -\flat^+_{-i-1} \flat_{-i},\quad
E_{-i}= \flat_{i+1}^+\flat_{i} -\flat^+_{-i} \flat_{-i-1}
\label{15.3}
\eeq
for $i=1,\, \cdots n-1$, and
\beq
E_n=\flat_{n-1}^+\flat_{-n} -\flat^+_{n} \flat_{-n+1},\quad
E_{-n}= \flat_{-n}^+\flat_{n-1} -\flat^+_{-n+1} \flat_{n}.
\label{15.4}
\eeq
The fundamental weights are
\begin{equation}
\vec \lambda_j= \sum_1^j \vec e_k,\quad k\leq n-2, \quad
\vec \lambda_{n-1}={1\over 2} \left (\sum_1^{n-1} \vec e_k-
\vec e_n\right ),\quad \vec \lambda_{n}= \sum_1^{n}
{\vec e_k\over 2}.
\label{15.6}
\end{equation}
As is well--known \cite{B}, the first $n-2$ \, fundamental representations
are of the same nature -- in contrast with the last two ones.
Their weight vectors have integer components as equations  (\ref{15.6})
show. They are immediately  realized in the Fock space of the fermionic
operators just introduced as follows:
the state
\beq
|\lambda_p > \equiv \flat^+_p \cdots \flat^+_1 |0>,
\quad 1\leq p\leq n-2,
\label{lambdap}
\eeq
satisfy the highest weight Eqs.(\ref{A14}).
The state $|0>$  is the usual  vacuum state with zero occupation numbers,
such that
\beq
\flat_\ell |0>, \quad -n \leq \ell \leq n.
\label{vac}
\eeq

How can we get the last two representations? The trick is to introduce
the operators
\bea
c_\ell \equiv  {1\over \sqrt{2}}(\flat_\ell + \flat^+_{-\ell}),
&\quad &
c^+_\ell \equiv {1\over \sqrt{2}}(\flat^+_\ell + \flat_{-\ell}),  \nnn
d_\ell\equiv {1\over i\sqrt{2}}(\flat_\ell - \flat^+_{-\ell}), &\quad &
d^+_\ell\equiv {1\over i \sqrt{2}}(\flat^+_\ell - \flat_{-\ell}),
\quad 1\leq \ell \leq n;
\label{flatpm}
\eea
which satisfy, according to Eq.(\ref{anticom}),
\bea
\bigl [ c_\ell, c_m\bigr ]_+=
\bigl [ c^+_\ell, c^+_m\bigr ]_+=& 0, &  \quad
\bigl [ c_\ell, c^+_m\bigr ]_+=\delta_{\ell, \, m}; \nnn
\bigl [ d_\ell, d_m\bigr ]_+=
\bigl [ d^+_\ell, d^+_m\bigr ]_+=&0, & \quad
\bigl [ d_\ell, d^+_m\bigr ]_+=\delta_{\ell, \, m}; \nnn
\bigl [ c_\ell, d_m\bigr ]_+=
\bigl [ c^+_\ell, d^+_m\bigr ]_+=& 0, &  \quad
\bigl [ c_\ell, d^+_m\bigr ]_+=0.
\label{anticom2}
\eea
These new operators give   us  two other realizations  of the
 algebra $D_n$  by introducing
$$
E^{(1/2)}_{\vec e_i+ \vec e_j}=c_i^+c^+_{j}, \quad
E^{(1/2)}_{-\vec e_i-\vec e_j} =c_jc_{i}, \quad
E^{(1/2)}_{\vec e_i-\vec e_j}=c_i^+c_{j}, \quad
E^{(1/2)}_{-\vec e_i+\vec  e_j} =c^+_jc_{i},
$$
\beq
h^{(1/2)}_i=H^{(1/2)}_i-H^{(1/2)}_{i+1},
\quad h^{(1/2)}_n=H^{(1/2)}_{n-1}+H^{(1/2)}_n, \quad
H^{(1/2)}_i=c_i^+c_i-{1\over 2},
\label{15.f1}
\eeq
$$
\wtE^{(1/2)}_{\vec e_i+\vec  e_j}=d_i^+d^+_{j}, \quad
\wtE^{(1/2)}_{-\vec e_i-\vec e_j} =d_jd_{i}, \quad
\wtE^{(1/2)}_{\vec e_i-\vec e_j}=d_i^+d_{j}, \quad
\wtE^{(1/2)}_{-\vec e_i+\vec e_j} =d^+_jd_{i},
$$
\beq
\wth^{(1/2)}_i=\wtH^{(1/2)}_i-\wtH^{(1/2)}_{i+1},
\quad \wth^{(1/2)}_n=\wtH^{(1/2)}_{n-1}+\wtH^{(1/2)}_n, \quad
\wtH^{(1/2)}_i=d_i^+d_i-{1\over 2}.
\label{15.g1}
\eeq
For later use we note that a straightforward  calculation gives
\beq
E_{\pm \vec e_i\pm \vec e_j}=E^{(1/2)}_{\pm \vec e_i\pm \vec e_j}+
\wtE^{(1/2)}_{\pm \vec e_i\pm \vec e_j}, \quad
H_i=H^{(1/2)}_i+\wtH^{(1/2)}_i.
\label{add}
\eeq
After this Bogolubov type  transformation, the new vacuum state noted
$|0>^{(1/2)}$
which is  annihilated by the
operators $c_\ell$ and $d_\ell$, is given by
\beq
|0>^{(1/2)}= \flat_{-n}^+ \cdots \flat_{-1}^+ |0>, \quad
c_\ell |0>^{(1/2)}=d_\ell |0>^{(1/2)}=0.
\label{fermvac}
\eeq
Using either of the above  two sets we can construct the last two
highest weight vectors. Consider, for instance the $c$-oscillators.
One easily verifies that the states
\beq
|\lambda_{n-1}>=c_{n-1}^+\cdots c_{1}^+|0>^{(1/2)},
\quad |\lambda_n >=c_{n}^+\cdots c_{1}^+|0>^{(1/2)},
\label{ll}
\eeq
satisfy the highest-weight equations
\beq
E^{(1/2)}_i |\lambda_p > =0, \quad
h^{(1/2)}_i |\lambda_p > =\delta_{i ,\, p}, \quad p=n-1, \> n.
\label{highf}
\eeq
The generators (\ref{15.2}), (\ref{15.5}) commute with the fermionic
number\footnote{In this formula and in the following, summations
from $-n$ to $n$ do not include zero}
\beq
{\cal N}_F\equiv \sum_{\ell=-n}^n \flat_\ell^+ \flat_\ell,
\label{fermn}
\eeq
and a representation with weight $\lambda_p$, with $p\leq n-2$
is realized in the space with a fixed number ($p$) of
 fermions. Thus we call it a
bosonic representation. On the contrary, the operators
of the realization (\ref{15.f1}) or Eq.(\ref{15.g1}) do not commute
with ${\cal N}_F$. We call them fermionic representations.

The Fock space we are considering allows us to realize every
fundamental representations in the same Hilbert space.
This is instrumental for the coming discussions since the Toda
equations and the corresponding infinitesimal Pl\"ucker formulae
do in fact connect these different representations, so that they will be
most naturally understood in the Fock space of the $\flat$ operators.
Moreover, this Fock space contains additional highest weight vectors
which will be very useful as well. First, the fermionic fundamental
representations are realized twice, since we may also use formulae
similar to Eq.(\ref{highf}), obtained after replacing $c$- by $d$-oscillators.
We shall denote these states by $|{\widetilde \lambda_{n-1}}>$ and
$|{\widetilde \lambda_{n}}>$ (c.f. e.g. \cite{B}).
Second, there are other
states analogous to (\ref{lambdap}).
They are given by
$\flat_{n-1}^+\cdots \flat_{1}^+|0>$,
$\flat_{n}^+\cdots \flat_{1}^+|0>$, and
$\flat_{-n}^+ \flat_{n-1}^+\cdots \flat_{1}^+|0>$.
 These are
highest-weight states since
it follows from (\ref{15.3}), and (\ref{15.4}) that they are annihilated by
$E_i$, for $i=1,\, \cdots ,\, n$.
The corresponding highest weights are given by
\bea
h_i\flat_{n-1}^+\cdots \flat_{1}^+|0>
&=&(\delta_{i,\, n-1}+ \delta_{i,\, n})
\flat_{n-1}^+\cdots \flat_{1}^+|0>,\nnn
h_i\flat_{n}^+\cdots \flat_{1}^+|0>&=&2\delta_{i,\, n}
\flat_{n}^+\cdots \flat_{1}^+|0>, \nnn
h_i\flat_{-n}^+\flat_{n-1}^+\cdots \flat_{1}^+|0>&=&2\delta_{i,\, n-1}
\flat_{n}^+\cdots \flat_{1}^+|0>.
\nonumber
\eea
Comparing with (\ref{15.6}), one sees that their weights are
$\lambda_{n-1}+\lambda_n$, $2\lambda_n$, and
$2\lambda_{n-1}$, respectively. Thus we  write
\beq
|\lambda_{n-1}+\lambda_n>=\flat_{n-1}^+\cdots \flat_{1}^+|0>,
\>
|2\lambda_n>=\flat_{n}^+\cdots \flat_{1}^+|0>,  \>
|2\lambda_{n-1}>=\flat_{-n}^+ \flat_{n-1}^+\cdots \flat_{1}^+|0>.
\label{15.43}
\eeq
The representations generated by  these highest-weight states
are irreducible, but not fundamental.
We shall see how they  fit in the general scheme of
Toda solutions, where   they come out
naturally   in the fermionic derivation of Toda solutions.
 Last we note  ---this will  be useful later on---
that there is another realization of the fundamental highest weight vectors
of the bosonic type. Indeed, it is easy to see that
\beq
\left [ H_i,\, \flat_k\flat_{-k}\right ]=0,
\label{c1}
\eeq
so that there is another highest weight state
\beq
|\lambda_{-p}>\equiv \flat^+_{-p-1} \flat^+_{-p-2} \cdots
\flat^+_{-n}\flat^+_{n}\cdots \flat^+_{1}|0>
\label{mm}
\eeq
with the same highest weight as $ | \lambda_{p}>$,
that is $\lambda_{p} =\lambda_{-p}$.
{}From the viewpoint of the fermionic operators, a transition  from
$|\lambda_{-p}>$ to $ |\lambda_{p}>$ is equivalent to  the exchange
of $\flat_j$ with  $\flat_{-j}^+$. Indeed, we have
\bea
\flat_j^+  |\lambda_{p}>&=&0, \quad  1\leq j\leq p,\quad
\flat_j  |\lambda_{p}>=0, \> j>p,\> \hbox{ or}\>  j\leq -1, \nnn
\flat_{-j}  |\lambda_{-p}>&=&0, \quad  1\leq  j\leq p,\quad
\flat_{-j}^+ | \lambda_{-p}>=0, \> j>p,\> \hbox{ or}\>  j\leq -1.
\nnn
\eea

\subsection{The target spaces associated with bosonic representations}
For any given highest-weight vector
$|\lambda_p>$, we split the Lie algebra $D_n$ into two parts:
\beq
D_n=\Gcc p  \uplus \Gcp p
\label{15.7}
\eeq
The one   called   $ \Gcc p $ leaves $|\lambda_p>$ invariant;
it  forms  a  Lie algebra. The symbol $\Gcp p $ denotes the
orthogonal complement.
The corresponding coset, denoted $ \Cco p $,
 is generated by exponentiating its  linear span.
The mathematical properties of these cosets are re-derived in
appendix A using the present  fermionic realization.
Next we describe the geometrical properties of these cosets.

\subsubsection{The coset space associated with $\lambda_1$}
 Following the method just described,
and according to appendix A,   this space
is param\-etri\-zed by\footnote{For $D_n$  we exponentiate the Cartan
generator $h_p$  separately so that the explicit formulae do not become
too complicated (see
appendix).}
\bea
\label{15.12}
e^{\kappa_1 h_1} e^{\Omega_1} \flat_1^+ | 0 >, \quad \hbox{with}\>
\> \Omega_1&=&\sum_{k=2}^{n} \left (  x_{k} E_{-\vec e_1+\vec e_k}+
x_{-k} E_{-\vec e_1-\vec e_k}\right ),  \nnn
<0|\flat_1
 e^{- \kappab_1 h_1} e^{- \Omegab_1}, \quad \hbox{with}\>
\> \Omegab_1&=&\sum_{k=2}^{n} \left (  \xb_{k} E_{\vec e_1+\vec e_k}+
\xb_{-k} E_{\vec e_1-\vec e_k}\right ),
\eea
where $\kappa_1$, $\kappab_1$ and $x_{\pm k}$, $\xb_{\pm k}$ are
group parameters that will
give a special parametrization of the coset.
After some computations, one derives that
\beq
e^{\Omega_1} \flat_1^+ e^{-\Omega_1} =
 \flat_1^+ + \sum_{k=2}^{n}\sum_{\epsilon=\pm 1}
x_{\epsilon k} \flat_{\epsilon k} ^+
-\left (\sum_{k=2}^{n} x_{k} x_{-k} \right ) \flat_{-1}^+,
\label{15.13}
\eeq
with similar equations for the $\kappab$ and $\xb$ coordinates.
It is convenient to write
\bea
e^{\kappa_1 h_1} e^{\Omega_1} \flat_1^+ | 0 >=
\sum_{-n\leq A \leq n,\> A\not=  0}
X^A \flat_A^+ |0>,
\nnn
<0|\flat_1 e^{-\kappab_1 h_1} e^{-\Omegab_1} =
\sum_{-n\leq A \leq n,\> A\not=  0}
<0|\flat_A\Xb^A;
\label{15.14}
\eea
where
\bea
X^{1}=e^{\kappa_1},\, \;
X^{-1}= -e^{-\kappa_1}\sum_{k=2}^{n} x_k x_{-k}, \;
X^{\pm 2} =e^{\mp \kappa_1} x_{\pm 2},\;
X^{\pm k} = x_{\pm k},\, k > 2;
\nnn
\Xb^{1}=e^{\kappab_1},\, \;
\Xb^{-1}= -e^{-\kappab_1}\sum_{k=2}^{n} \xb_k \xb_{-k}, \;
\Xb^{\pm 2} =e^{\mp \kappab_1} \xb_{\pm 2},\;
\Xb^{\pm k} = \xb_{\pm k},\, k > 2.
\label{15.15}
\eea
The functions $X^A$, $\Xb^A$  satisfy the quadratic equations
\begin{equation}
\sum_{A} X^{A} X^{-A}=0, \quad
\sum_{A} \Xb^{A} \Xb^{-A}=0.
\label{15.16}
\end{equation}
In this equation and in the following, the sums over $A$ run from
$-n$ to $n$, with $0$ excluded.
 It is convenient to introduce the
following notation
\beq
X\equiv \sum_A X^A \flat^+_A,\>
\bar X \equiv \sum_A \bar X^A \flat_A.
\label{15.17}
\eeq
It is natural to define a K\"ahler metric on $\Cco 1 $ derived from the
  K\"ahler potential
\beq
{\cal K}^{(1)}(X,\Xb)\equiv
\ln \left [ <0|\flat_1  e^{-\Omegab_1}e^{-\kappab_1 h_1}
e^{\kappa_1 h_1} e^{\Omega_1} \flat_1^+ | 0 > \right ]=
\ln \left [ < 0 | \Xb X |0> \right ].
\label{15.18}
\eeq
which  has an obvious group invariance.
Together with condition (\ref{15.16}), this  completes  the definition
of the manifold  $\Cco 1 $.
 It  may be understood as a submanifold of
$CP^{2n-1}$.  Indeed,  the K\"ahler potential
coincides with the one of Fubini-Study, and  the
quadratic constraints are invariant under the rescalings
\beq
X^A\to X^A \rho (X), \quad
\Xb^A\to \Xb^A \rhob (\Xb),
\label{15.20}
\eeq
that leave the points  of  $CP^{2n-1}$ invariant.
The manifold $\Cco 1 $ is thus a quadric
in $CP^{2n-1}$.

 Choose coordinates such that
$X^1=\Xb^1=1$. Then we can solve the constraint (\ref{15.16}),
obtaining
\beq
X^{-1}=-\sum_{A=2}^n X^A X^{-A}, \quad
\Xb^{-1}=-\sum_{A=2}^n \Xb^A \Xb^{-A};
\label{15.21}
\eeq
and the K\"ahler potential becomes
\beq
{\cal K}^{(1)}(X,\Xb)=
\ln \left [ 1+\sum_{A=2}^n \left (X^A \Xb^{A}
+X^{-A} \Xb^{-A}\right ) +\left (\sum_{B=2}^n X^B X^{-B}\right )
\left (\sum_{\Bb=2}^n \Xb^\Bb \Xb^{-\Bb}\right ) \right ]
\label{15.22}
\eeq
This  is the equivalent of the Fubini-Study metric for the present case.
\subsubsection{The coset space associated with generic bosonic
fundamental highest weight.  }

The discussion is very close to the above. The manifold $\Cco p $ is
parametrized by
\bea
e^{\kappa_p h_p} e^{\Omega_p} |\lambda_p>
&=&\sum_{A_1,\, \cdots ,  \,A_p}
X^{[p] \,p,\, A_p}  \cdots   X^{[p]\, 1,\, A_1}
\flat_{A_p}^+ \cdots \flat_{A_1}^+ |0>,  \nnn
< \lambda_p | e^{-\Omegab_p} e^{-\kappab_p h_p}
&=& \sum_{A_1,\, \cdots ,  \,A_p}
<0| \flat_{A_1} \cdots  \flat_{A_p} \Xb^{[p] \,p,\, A_p}
 \cdots   \Xb^{[p]\, 1,\, A_1},
\label{par}
\eea
where we have let
\bea
\sum _A X^{[p]\, \alpha,\, A} \flat^+_A & \equiv& X^{[p]\, \alpha}
=e^{\kappa_p h_p} e^{\Omega_p} \flat_\alpha^+
e^{-\kappa_p h_p} e^{-\Omega_p},
\nnn
\sum _A \Xb^{[p]\, \alpha,\, A} \flat^+_A & \equiv& \Xb^{[p]\, \alpha}
=e^{-\kappab_p h_p} e^{-\Omegab_p} \flat_\alpha e^{\kappab_p h_p} e^{\Omegab_p}
\label{coo}
\eea
The natural   K\"ahler potential
\beq
{\cal K}^{[p] }\equiv \ln \left [
< \lambda_p | e^{-\Omegab_p} e^{-\kappab_p h_p}
e^{\kappa_p h_p} e^{\Omega_p} |\lambda_p> \right ]
\label{kpp}
\eeq
 takes the form of Kobayashi-Nomizu
\beq
{\cal K}^{[p] }(X^{[p]},\Xb^{[p]})=
\ln \left [ < 0 | \Xb^{[p]  1}\cdots \Xb^{[p] p-1 }\Xb^{[p] p}
X^{[p]  p}X^{[p] p-1}\cdots X^{[p] 1} |0> \right ].
\label{15.31}
\eeq
The precise connection between the coordinates $X$, $\Xb$ and the group
parameters is given in appendix, where it is also shown that the
coordinates $X$, and  $\Xb$ satisfy the quadratic relations.
\beq
  \left \{  \begin{array}{c}
\sum_{A} X^{[p] \, \alpha,\, A} X^{[p]\, \beta,\, -A}=0, \nnn
\sum_{A} \Xb^{[p] \, \alpha,\, A} \Xb^{[p]\, \beta,\, -A}=0;
\end{array} \right.
\quad 1\leq \alpha \leq p,\>
\quad 1\leq \beta  \leq p.
\label{15.29}
\eeq
The origin of these
relations is that  there are less
group parameters than coordinates $X$ and $\Xb$.
A  compact proof of these identities  goes as
follows\footnote{We treat the  coordinates $X$.
The  coordinates $\Xb$ could
obviously be discussed   similarly.}.   Due to the special
form of the generators of $D_n$, which is
displayed in (\ref{15.2})--(\ref{15.5}), there exists  a
symmetry  which we call charge conjugation, and
denote by a superscript $c$. It is defined by
\beq
\left ( \flat_{A_k}^+\cdots \flat_{A_1}^+
\flat_{B_\ell}\cdots \flat_{B_1}\right )^c
=
\left (\flat_{-A_k}^+\cdots \flat_{-A_1}^+
\flat_{-B_\ell}\cdots \flat_{-B_1}\right )^+,
\label{15.55}
\eeq
and relations (\ref{15.2})--(\ref{15.5}) show that
\beq
E^c_{\pm \vec e_i\mp \vec e_j} = -E_{\pm \vec e_i\mp \vec e_j}, \>
h^c_i = -h_i,
\label{15.56}
\eeq
The origin of this charge conjugation is the orthogonality of $D_n$.
It  transforms
the first line of (\ref{coo}) into
\[
\sum _A X^{[p]\, \alpha,\, A} \flat^+_{-A}
=e^{\kappa_p h_p} e^{\Omega_p} \flat_{-\alpha} e^{-\kappa_p h_p} e^{-\Omega_p},
\]
so that
\[
<0| \flat_{-\alpha} \flat^+_{\beta}|0> =
 <0| e^{\kappa_p h_p} e^{\Omega_p}\flat_{-\alpha} e^{-\Omega_p}
e^{-\kappa_p h_p} e^{\xi_p h_p}e^{\Omega_p}
\flat^+_{\beta}e^{\Omega_p}|0>=
\sum_{A} X^{[p] \, \alpha,\, A} X^{[p]\, \beta,\, -A}.
\]
Since, obviously $<0| \flat_{-\alpha} \flat^+_{\beta}|0>=0$,
 relations (\ref{15.29}) follows.
This fact   completes\footnote{Obviously, the coordinates
$X^{[p]\, \alpha,\, A}$ and $\Xb^{[p]\, \alpha,\, A}$ satisfy, in
addition,  Pl\"ucker type relations similar to (\ref{A26}).
We shall not dwell on this aspect.} the definition of the
 manifold $\Cco p $.
Next we show that it is
a submanifold of the
usual Grassmannian manifold $Gr(2n-1| p)$. In general, $Gr(2n-1| p)$
is the set of $(2n-1) \times p$ matrices ${\cal F}_p$
 with the equivalence relation ${\cal F}_p \sim \rho  {\cal F}_p$,
where  $\rho$ is  an arbitrary $p\times p$ matrix,
that is the generalization
of (\ref{15.20}), which  corresponds to $p=1$.  The geometrical meaning
of this equivalence is well known: given ${\cal F}_p$,
one defines hyperplanes in $CP^{2n-1}$ by equations of the form
$Z^A(t)= \sum_\alpha {\cal F}_p^{\alpha,\, A}t_\alpha$. The
equivalence relations is equivalent to linear transformations of
the parameters $t_\alpha$. Thus the Grassmannian  describes geometrical
hyperplane which should  not depend upon their parametrizations.
In our case the coordinates are $X^{[p]\, \alpha,\, A}$, and
$\Xb^{[p]\, \alpha,\, \Ab}$. It is well known that the metric
derived from the K\"ahler potential
 (\ref{15.31}) is invariant under  the transformation
\beq
X^{[p]  \alpha,\, A}\to \rho(X^{[p]})_\beta^\alpha
\> X^{[p] \, \beta,\, A},\>
\Xb^{[p]  \alpha,\, \Ab}\to \rhob(\Xb^{[p]})_\beta^\alpha
\> \Xb^{[p] \, \beta,\, \Ab}.
\label{15.32}
\eeq
Moreover, it  is easy to
see that, if $X^{[p] \, \alpha,\, A}$ satisfies condition
Eq.\ref{15.29}, this is also true for $\sum_{\beta} \rho_\beta^\alpha
X^{[p] \, \beta,\, A}$.     Thus these  conditions  define
a quadric in the Grassmannian $Gr(2n-1| p)$.

\subsubsection{The three  additional coset spaces}
\label{additional coset}

It is obvious that the previous description of the cosets
extends to the representations with the highest vectors $|\lambda_{n-1}+
\lambda_n>$,
$|2\lambda_n>$, and $|2\lambda_{n-1}>$ (see definitions (\ref{15.43}))
without any problem. The first two cases are direct extensions of the
formulae given for $|\lambda_p>$ with $p\leq n-1$.  The last one is obtained
from the calculation for $|2 \lambda_{n-1}> $, by exchanging everywhere
$\flat_n$ with $\flat_{-n}$. Some details are given in appendix.
Always using  similar
notations, we  introduce
\beq
X^{[\lambda]  r} =e^{\kappa_\lambda h_\lambda} e^{\Omega_\lambda} \flat_r^+
e^{-\Omega_\lambda} e^{-\kappa_\lambda h_\lambda},
\Xb^{[\lambda]  r} =e^{\kappab_\lambda h_\lambda} e^{\Omegab_\lambda} \flat_r
e^{-\Omegab_\lambda} e^{-\kappab_\lambda h_\lambda},
\label{15.333}
\eeq
\beq
{\cal K}^{[\lambda] }(X^{[\lambda]},\Xb^{[\lambda]})=
\ln \left [ < 0 | \Xb^{[\lambda]  1}\cdots \Xb^{[\lambda] p_\lambda}
X^{[\lambda]  p_\lambda}\cdots X^{[\lambda] 1} |0> \right ].
\label{15.31g}
\eeq
The same reasoning as above shows that one has the quadratic conditions
\beq
\sum_A X^{[\lambda]  p\, A} X^{[\lambda]  q\, -A}=0.
\label{15.31h}
\eeq
These K\"ahler potentials take the Kobayashi-Nomizu form. They will
appear naturally in connection with the explicit solution of the
$D_n$-Toda equations.  The formulae just given define manifolds
$\Cco {\lambda}  $ for $\lambda=\lambda_{n-1}+\lambda_n$,
 $\lambda=2\lambda_{n}$, and  $\lambda=2\lambda_{n-1}$.

\subsection{The case of the  two fermionic fundamental representations}
As already recalled,
these two fundamental representations are of a
completely different nature \cite{B}.
While the $n-2$ first ones have dimensions
${2n \choose p}$, $p\leq n-2$, they have
dimension $2^{n-1}$. We make use of the realization (\ref{15.f1});
the highest weight states
are given by (\ref{ll}). This being established, the discussion proceeds
exactly as before. The coordinates are given by
\bea
\begin{array}{ccc}
e^{\kappa_p  h^{(1/2)}_p} e^{\Omega^{(1/2)}_p} |\lambda_p>&
=&\Xs^{[p]\, p} \Xs^{[p]\, p-1} \cdots \Xs^{[p]\, 1} |0>, \nnn
\Xs^{[p]\,\alpha } \equiv
e^{\kappa_p  h^{(1/2)}_p}e^{\Omega^{(1/2)}_p} c^+_\alpha
e^{-\Omega^{(1/2)}_p}e^{-\kappa_p  h^{(1/2)}_p}&=&
 \sum_{A>0} \left (X^{[p]\,\alpha, \, A } c_A^+
+X^{[p]\,\alpha, \, -A } c_A\right ), \end{array}
\nnn
 \begin{array}{ccc}
< \lambda_p| e^{-\kappab_p  h^{(1/2)}_p} e^{-\Omegab^{(1/2)}_p} &=&
<0| \Xsb^{[p]\, p} \Xsb^{[p]\, p-1} \cdots \Xsb^{[p]\, 1}, \nnn
\Xsb^{[p]\,\alpha } \equiv
e^{\kappab_p  h^{(1/2)}_p}e^{\Omegab^{(1/2)}_p} c_\alpha
e^{-\Omegab^{(1/2)}_p}e^{-\kappab_p  h^{(1/2)}_p}&=&
 \sum_{A>0} \left (\Xb^{[p]\,\alpha, \, A } c_A
+\Xb^{[p]\,\alpha, \, -A } c^+_A\right ), \end{array}
\nnn
\label{15.h}
\eea
for $\alpha \leq p$, $p=n-1$, and $p=n$.
As before
there are again quadratic constraints which may be derived by writing:
\[
e^{\kappa_p  h^{(1/2)}_p}e^{\Omega^{(1/2)}_p}
\left [ c_\alpha^+,  c_\beta^+\right ]_+
e^{-\Omega^{(1/2)}_p}e^{-\kappa_p  h^{(1/2)}_p}
=0= \left [ \Xs^{[p]\,\alpha }, \Xs^{[p]\,\beta }
\right ].
\]
This gives again  relations (\ref{15.29}), now with $p=n-1$ and $p=n$.
The natural K\"ahler potential, that is (\ref{kpp})  for $p=n-1$, $n$,
is given by
\beq
{\cal K}^{[p] }(X^{[p]},\Xb^{[p]})=
\ln \left [ < 0 | \Xsb^{[p]  1}\cdots \Xsb^{[p] p-1 }\Xsb^{[p] p}
\Xs^{[p]  p}\Xs^{[p] p-1}\cdots \Xs^{[p] 1} |0> \right ].
\label{15.41}
\eeq
A priori it is different from the Kobayashi-Nomizu form,
since the operators
$\Xs$, and $\Xsb$ involve both creation and annihilation operators.
We shall spell out the connection below.
This completes\footnote{Here also there are additional
quadratic relations similar to (\ref{A26}).}
 the definition of  the manifolds $\Cco p $, for
$p=n-1$, $n$.

\subsection{Connection between K\"ahler potentials}

The last three coset spaces just discussed are not associated with
fundamental representations. We now show that they  can be re-expressed
in terms of the potential associated with the last two fundamental
highest weights. This, of course, is due to the fact that their highest
weights are linear combinations of $\lambda_{n-1}$ and $\lambda_n$.
The present fermionic method gives a quick derivation of this fact. Indeed,
we already mentioned that the fermionic fundamental representations are
realized twice, once in terms of the $c$-oscillators (\ref{15.f1}), and
once in terms of the $d$-oscillators (\ref{15.g1}). Using  formulas
(\ref{flatpm}), one sees that
\bea
c_n^+\cdots c_1^+ d_n^+ \cdots d_1^+ |0>^{(1/2)} &=&
i^n (-1)^{n(n+1)/2} |2\lambda_n>, \nnn
c_{n-1}^+\cdots c_1^+ d_n^+ \cdots d_1^+ |0>^{(1/2)} &=&
\begin{array}{l} i^n (-1)^{n(n+1)/2}/  \sqrt 2 \times \nnn
 ( |\lambda_{n-1}+ \lambda_n>+
|  \lambda_{-(n-1)}+  \lambda_{-n}>),
 \end{array}   \label{zut}\\
c_{n-1}^+\cdots c_1^+ d_{n-1}^+ \cdots d_1^+ |0>^{(1/2)} &=&
i^{n-1} (-1)^{n(n-1)/2} |2\lambda_{n-1}>,
\label{r3}
\eea
where the state $|\lambda_{-(n-1)}+  \lambda_{-n}>$ is defined
by the obvious generalization of (\ref{mm}). The formulae just written
are clearly consistent with (\ref{add}). Now, we may re-derive the
expressions (\ref{15.31g}) of the K\"ahler potentials ${\cal K}^{[\lambda]}$
with $\lambda=\lambda_{n-1}+\lambda_n$,
$\lambda=2\lambda_n$, and $\lambda=2\lambda_{n-1}$, using the l.h.s.
of the last equations  together with (\ref{15.g1}). In this way, the
calculations involving the $c$ and $d$ operators  become completely
separated. Each of them is entirely specified by the group properties
of the fermionic fundamental representations which do not depend upon
 of the realization chosen. Moreover, the dimensions of the
manifolds  involved coincide, so that there are   natural
mappings  between them.
 It is then easy to conclude that the
K\"ahler potentials are related by
\bea
{\cal K}^{[2\lambda_n] }(X^{[n]},\Xb^{[n]}) &=&
2 {\cal K}^{[n] }(X^{[n]},\Xb^{[n]}),\nnn
{\cal K}^{[\lambda_{n-1}+\lambda_{n-1}] }(X^{[n]},\Xb^{[n]})  &=&
 {\cal K}^{[n] }(X^{[n]},\Xb^{[n]})
+{\cal K}^{[n-1] }(X^{[n]},\Xb^{[n]}), \nnn
{\cal K}^{[2\lambda_{n-1}] }(X^{[n-1]},\Xb^{[n-1]}) &=&
2 {\cal K}^{[n-1] }(X^{[n-1]},\Xb^{[n-1]}).
\label{15.c5}
\eea
These  relations will be important later on.

\section{Generalized Pl\"ucker embeddings for $D_n$}
\subsection{Definitions}

Let us introduce the following definitions which will be motivated
by the forthcoming discussions.
\begin{definition}{$D_n$--W-surfaces.}
\label{W-surfaces}

The W surfaces associated with the Lie algebra
$D_n$ are  two dimensional surfaces $\Sigma^{[1]}$
in $\Cco 1 $ defined by the equations
\beq
X^{[1]A}=f^A(z), \quad \Xb^{[1]A}=\fb^A(\zb);
\label{wsurf}
\eeq
where
$f^A(z)$ and $\fb^A(\zb)$   satisfy the quadratic
differential relations
\[
\sum_{A>0} f^{(a)\, A}(z)\> f^{(b)\, -A}(z)=\delta_{a, n-1} \delta_{b, n-1},
\quad \sum_{A>0} \fb^{(a)\, A}(\zb)\> \fb^{(b)\, -A}(\zb)
= \delta_{a, n-1} \delta_{b, n-1},
\]
\beq
\hbox{for }\>  0\leq a,\, b \leq n-1.
\label{15.61}
\eeq
\end{definition}
In the last formula, and hereafter, upper indices in between round
parentheses denote the order of derivatives in $z$ or $\zb$.
For $p=q=1$ the conditions just written are, of course, necessary for
$\Sigma^{[1]}$ to be a submanifold of $\Cco 1 $ (see (\ref{15.16})).
The additional conditions
will be needed for consistency with the following
\begin{definition}{Associated surfaces.}
\label{Associated surfaces}

Given  any $D_n$--W-surfaces, in the sense of  definition
\ref{W-surfaces}, it is convenient to  introduce  a family
of surfaces $\Sigma^{[p]}$
in $\Cco p $, $p=2$, $\cdots$, $n$ defined by the equations
\beq
X^{[p]A}=f^{(p-1)A}(z)-\delta_{p, n} f_{\parallel}^{[n-1]A}(z), \quad
\Xb^{[p]A}=\fb^{(p-1)A}(\zb) -\delta_{p, n} \fb_{\parallel}^{[n-1]A}(\zb),
\label{ass}
\eeq
where
\bea
f_{\parallel}^{[n-1]A} = \sum_q\left
( \Theta^{-1}\right )_{n-1\>  q} f^{(q) A}, \> A>0,
\quad f_{\parallel}^{[n-1]A}=0,\quad A<0, \nnn
\fb_{\parallel}^{[n-1]A} = \sum_q\left
( \bar \Theta^{-1}\right )_{n-1\>  q} \fb^{(q) A}, \> A>0,
\quad \fb_{\parallel}^{[n-1]A}=0,\quad A<0;
\label{fpar}
\eea
\beq
\Theta_{p q}\equiv \sum_{A>0} f^{(p) A} f^{(q) -A},\quad
\bar \Theta_{p q}\equiv \sum_{A>0} \fb^{(p) A} \fb^{(q) -A}; \quad
\quad 0\leq p, \> q \leq n-1.
\label{theta}
\eeq
\end{definition}
The
definition of $f_{\parallel}$ and $\fb_{\parallel}$ is
such that
\beq
\sum_A f_{\parallel}^{[n-1]A} f^{[p] -A}=
\sum_A \fb_{\parallel}^{[n-1]A} \fb^{[p] -A}= \delta_{p, n-1}.
\label{17.20}
\eeq
In view of relations (\ref{15.61}), it follows that (\ref{ass}) are
  compatible with
conditions (\ref{15.29}), and $\Sigma^{[p]} \in
\Cco p $, as the definition claims.   As usual, the geometrical interpretation
of (\ref{ass}) should be  that a point of $\Sigma^{[p]}$ represents the
osculating hyperplane with contact of order $p-1$ at the point
$ X^{[1]A}=f^A(z)$, $\Xb^{[1]A}=\fb^A(\zb)$ of $\Sigma^{[1]}$.
Conditions (\ref{15.61}) precisely ensure that $\Sigma^{[p]}$ has such
a contact with the quadric of equations  $\sum_A X^A X^{-A}=0$,
$\sum_A \Xb^A \Xb^{-A}=0$, which defines $\Sigma^{[1]}$ as a submanifold
of $CP^{2n-1}$. Thus we shall consider the definition just given
as the one of the generalized Pl\"ucker embedding associated with $D_n$.
The above definition makes sense at generic points of the W-surface
where $\Theta$ and $\bar \Theta$ are invertible matrices.

\subsection{Pl\"ucker embedding from  Toda dynamics}

The main aim of the present subsection is the derivation of the following
\begin{theorem}
\label{TodaPlucker}

Associated with
any solution of the $D_n$-Toda equations, there exist  a $D_n$--W-surface
and a family of associated  surfaces as introduced  by definitions
\ref{W-surfaces} and \ref{Associated surfaces},
where  $f$ and $\fb$ are given by
\beq
f^{A} =e^{-\xi_1} F^A,\quad \fb^{A} =e^{-\xib_1} \Fb^{A} .
\label{15.54}
\eeq
\[
F^{ 1}=1,\> F^{ 2}=(1),\>
F^{ k}=(1,2,\cdots ,k-1),\, k\leq n;
\]
\[
F^{ -n}=(1,2,\cdots ,n-2,n),\>
\]\[
F^{ -n+1}=-(1,2,\cdots ,n-2,n-1,n)-(1,2,\cdots ,n-2,n,n-1)
\]\[
F^{ -\ell}=
(-1)^{n-\ell}\Bigl [\> (1,2,\cdots ,n-2,n-1,n,n-2,\cdots ,\ell)+
\]
\beq
(1,2,\cdots ,n-2,n,n-1,n-2,\cdots ,\ell)\>\Bigr],\>  \ell <n-1.
\label{15.53}
\eeq
The last equation uses the following  compact notation for the repeated
integrals over screening functions:
\beq
(i_1,i_2,\cdots ,i_r)\equiv
\int_{z_0}^zdx_1s_{i_1}(x_1)\int_{z_0}^{x_1}dx_2s_{i_2}(x_2)
\cdots \int_{z_0}^{x_{r-1}}dx_{r}s_{i_r}(x_r),
\label{15.50}
\eeq
and the anti-holomorphic parts are given by similar expressions.
\end{theorem}

\proof  We have to show  that the functions  $f^A$ defined by formulas
(\ref{15.54})
obey conditions (\ref{15.61}). Using the explicit realization (\ref{15.5}),
(\ref{15.3}), it is easy to
verify
that
\beq
M(z) \flat_1^+ M(z)^{-1}=\sum_A F^{ A}(z)\>  \flat_A^+\equiv F(z),
\label{15.52}
\eeq
where $F^A$ is given by (\ref{15.53}).
According to (\ref{15.45} one has,
\beq
d(M(z) \flat_{k}^+ M^{-1}(z)) /dz =M(z)
[L, \, \flat_{k}^+] M^{-1}(z), \quad
L=\sum_i s_i E_{-i}.
\label{17.3}
\eeq
Together  with Eq.\ref{15.3} this gives for $1\leq k\leq n-1$,
\beq
M(z) \flat_{k}^+ M^{-1}(z) = D_k F,
\label{17.4}
\eeq
where we introduced the notation
\beq
D_k\equiv
{1\over s_{k-1}}{d\over dz}
{1\over s_{k-2}}{d\over dz} \cdots {1\over s_{1}}{d\over dz},
\quad  k \geq 2; \quad D_1=1.
\label{17.4x}
\eeq
By the charge conjugation (\ref{15.55}),
the differential equation for
$M(z)$ becomes $ d M^c/ dz$ $=-(\sum_{j=1}^n s_j(z)E_{-j})M^c$,
so that
$M^c=M^{-1}$. This was expected since we are dealing with  the
orthogonal algebra. After charge conjugation, Eq.\ref{17.4} becomes
\beq
M(z) \flat_{-k} M(z)^{-1}= D_k F^c, \> \hbox{where} \>
F^c\equiv \sum_A F^A \flat_{-A}.
\label{15.57}
\eeq
The method for  deriving    quadratic relations is  to  consider
\[
<0|M(z) \flat_{-k}M^{-1}(z)M(z)\flat^+_\ell M^{-1}(z)|0>=
<0|\flat_{-k}\flat^+_\ell |0>=0.
\]
One  re-expresses
the l.h.s. using (\ref{17.4}) and (\ref{15.57}). This gives
\beq
\sum_A D_k F^{ A}(z)\> D_\ell F^{ -A}(z)=0,\quad  1\leq k < \ell  \leq n-1.
\label{15.58}
\eeq
It follows from (\ref{15.54}) that
\bea
D_k F^A&=& e^{\xi_k-\xi_{k-1}}  f^{(k-1)\, A}
+ \hbox{ lower order derivatives}, \quad k \leq n-2; \nnn
D_{n-1} F^A&=& e^{ \xi_n+\xi_{n-1}-\xi_{n-2}}  f^{(n-2)\, A}
+ \hbox{ lower order derivatives}.
\label{df}
\eea
Combining the last two formulae, one concludes
\beq
\sum_A f^{(a)\, A}(z)\> f^{(b)\, -A}(z)=0
\quad \hbox{for } \> 0\leq a,\, b \leq n-2.
\label{qr1}
\eeq
The case $k=n-1$ is different, since (\ref{17.3}) gives
\bea
D_n F&=& M(z) \flat_{n}^+ M^{-1}(z) +{s_n\over s_{n-1}}
M(z) \flat_{-n}^+ M^{-1}(z), \nnn
D_n F^c&=&  M(z) \flat_{-n} M^{-1}(z) +{s_n\over s_{n-1}}
M(z) \flat_{n} M^{-1}(z);
\label{eqn}
\eea
\[
<0| D_k F^c D_n F |0> =2  \delta_{n, k}  s_n/ s_{n-1}.
\]
Making use of (\ref{df}), together with the equation
\beq
D_n F^A= e^{\xi_n-\xi_{n-1}}  f^{(n-1)\, A}
+ \hbox{ lower order derivatives},
\label{nnn}
\eeq
one obtains
\[
\sum_{A>0}  f^{(a)\, A}(z)\> f^{(n-1)\, -A}(z)=
\delta_{a, n-1}, \quad 0\leq k \leq n-1.
\]
This completes the proof that (\ref{15.54}) and
(\ref{15.53}) define embedding fuctions $f$ that obey
conditions  (\ref{15.61}). The case of $\fb$ is similar.
\qed

A direct  consequence of the fermionic method we are using is the
\begin{corollary}{K\"ahler potentials  from Toda fields.}
\label{KahlerToda}

The intrinsic metric  of the surface $\Sigma^{[p]}$ defined by the
above theorem, is derivable from  the K\"ahler potential equal to
$-\Phi_p$, $p=1$, $\cdots$, $n$.
\end{corollary}

\proof Consider, first, the  representation with highest-weight vector
$\flat_p^+\cdots \flat_1^+|0>$  for $1\leq p \leq n-2$.
Making use of (\ref{17.4}),
one concludes that
\beq
M(z) \flat_p^+ \cdots \flat_1^+|0> =
\prod_{r=1}^{p-1}s_r^{p-r} \sum_{A_{p-1}}F^{(p-1)\, A_{p-1}}
\flat_{A_{p-1}}^+
\cdots \sum_{A_{1}}F^{ A_{1}} \flat_{A_1}^+|0>.
\label{15.63}
\eeq
Is is well--known that the inverse of the Cartan matrix is expressible
in terms of the fundamental weights; $(K^{(D_n)\,-1})_{ij}=
\vec \lambda_i. \vec \lambda_j$. Using
(\ref{15.6}), and substituting (\ref{15.54})
for $f^A$,  one finds finally that
\bea
e^{-\xi_p} M(z) \flat_p^+ \cdots \flat_1^+|0> & = & f^{(p-1)} \cdots
f^{(1)} f|0>, \nnn
e^{-\xib_p}  <0| \flat_1\cdots \flat_p \Mb^{-1}(\zb) & = & <0|
\fb \fb^{(1)}\cdots  \fb^{(p-1)}.
\label{xxx}
\eea
According to (\ref{A3}), (\ref{15.45}), and (\ref{15.31}), this gives
the desired relation
\beq
{\cal K}^{[p]}( f, \cdots ,  f^{(p-1) },
\fb, \cdots , \fb^{(p-1) })
= -\Phi_{p}(z,\zb).
\label{17.17x}
\eeq
Next we consider the associated surfaces  in $\Cco p $,
for $p\geq n-2$.
Of course, this part makes use of the fermionic
realization in terms of the $c$-fermionic operators (see (\ref{15.f1})).
A calculation similar to the one that leads to (\ref{15.52})
gives
\beq
M(z) c_{1}^+ M(z)^{-1}=\sum_{A >0}
\left (F^{ A}(z)\>  c^+_{A} + F^{-A}(z)\>  c_{A}\right
)
\equiv \Fs(z).
\label{17.2}
\eeq
Since one has,
$$d(M(z) c_{k}^+ M^{-1}(z)) /dz =M(z)
[{\cal E}, \, c_{k}^+] M^{-1}(z),
{\cal E}=\sum_i s_i E_{-i},
$$
it follows that, for $k\leq n-2$,
\beq
M(z) c_{k}^+ M^{-1}(z) = D_k \Fs.
\label{17.4s}
\eeq
In agreement with (\ref{15.54}), we let
\beq
\fs (z) \equiv e^{-\xi_1}\Fs (z),
\label{17.5}
\eeq
obtaining
\beq
D_k \Fs  =
e^{\xi_k-\xi_{k-1}} \fs ^{(k-1)} +\hbox{lower derivative terms},
\quad \hbox { for } k\leq n-2,
\label{17.6}
\eeq
and, thereof,
\beq
M(z) c^+_{k} \cdots       c_1^+ |0>^{(1/2)}
=e^{\xi_k} \fs^{(k-1)} \cdots \fs^{(1)} \fs |0>^{(1/2)},
\quad \hbox {for } k\leq n-2.
\label{17.7}
\eeq
So far, this is much like what was discussed in the previous
case.  For $k=n-1$,
 the calculation is again similar, but the expression of
$\lambda_{n-1}$ is   different, and one finds
$$M(z) c_{n-1}^+ M^{-1}(z) = D_{n-1} \Fs =
e^{\xi_n+\xi_{n-1}-\xi_{n-2}} \fs ^{(n-2)}+\hbox{lower derivative terms};
$$
\beq
M(z) c^+_{n-1} \cdots  c_1^+ |0>^{(1/2)}
=e^{\xi_n+\xi_{n-1}}  \fs^{(n-2)} \cdots \fs^{(1)} \fs|0>^{(1/2)}.
\label{17.8}
\eeq
The orthogonality conditions (\ref{15.61})
 can be re-derived using the $c$-operators. They come out very simply
from the obvious relations
\beq
\Bigl [ M(z) c^+_{k} M^{-1}(z), \>
M(z) c^+_{\ell } M^{-1}(z) \Bigr ]_+=0,
\label{17.9}
\eeq
and from the counterpart of (\ref{eqn}), that is,
$$d(M(z) c_{n-1}^+ M^{-1}(z)) /dz =s_{n-1} M
c_{n}^+ M^{-1} +s_{n} M
c_{n} M^{-1}.
$$
This gives the equation
\beq
M c^+_{n} M^{-1} =
D_n \Fs  -{s_n\over s_{n-1}} M c_{n} M^{-1}.
\label{17.11}
\eeq
that will be useful below.
Next consider the case of the associated surface in $\Cco {n-1}  $.
The
 embedding is very similar
to the case $p\leq n-2$, since
one makes only use of conditions  (\ref{15.61})  for $p \leq n-2$, and
$q\leq n-2$,  which are homogeneous. Note, however, that formula (\ref{17.8})
involves the factor $\exp(\xi_n+\xi_{n-1})$, instead of the factor
$\exp(\xi_{n-1})$ that would be the direct generalization of
the
bosonic representation case. Thus one finds
\beq
{\cal K}^{[n-1]}( f, \cdots ,  f^{(n-2) },
\fb, \cdots , \fb^{(n-2) })
= -(\Phi_{n-1}(z,\zb) +\xi_n +\xib_n).
\label{17.17}
\eeq
Finally let us discuss the associated surface in $\Cco {n}  $.
According to (\ref{17.11}),
$$
\Bigl [ M(z) c^+_{k} M^{-1}(z), \>
M(z) c^+_{n } M^{-1}(z) \Bigr ]_+=
0=
$$
$$
\Bigl [ D_n \Fs  -{s_n\over s_{n-1}} M c_{n} M^{-1},
D_n \Fs  -{s_n\over s_{n-1}} M c_{n} M^{-1} \Bigr ]_+.
$$
Thus, by keeping the second term, one   arrives  at an homogeneous
relation. Comparing with Eqs.\ref{17.20}, one concludes that
\beq
M c_{n} M^{-1}\equiv \fs_{\parallel}^{[n-1]}
=\sum_{A >0}  f_{\parallel}^{[n-1] A } c_A.
\label{17.19}
\eeq
The fact that the last formula involves only annihilation
operators is a direct consequence of the explicit realization (\ref{15.f1}).
It is easily seen that the second term of
(\ref{17.11})  drops out when one computes the
generalization of (\ref{17.7}) for $k=n$. One gets
\beq
M c^+_{n} \cdots  c_1^+ |0>^{(1/2)}
=e^{2\xi_n}  \fs^{(n-1)} \cdots \fs^{(1)} \fs |0>^{(1/2)}.
\label{17.12}
\eeq
Thus we have
$$
{\cal K}^{[n]}( f, \cdots ,  f^{(n-2)},  f^{(n-1) }-f_{\parallel}^{[n-1]} ,
\fb, \cdots , \fb^{(n-2) } \fb^{(n-1) }-\fb_{\parallel}^{[n-1]})=
$$
\beq
{\cal K}^{[n]}( f, \cdots ,  f^{(n-1)},
\fb, \cdots , \fb^{(n-1) }) =
 -(\Phi_{n}(z,\zb) +\xi_n +\xib_n).
\label{17.22}
\eeq
The outcome of the preceding discussion
is that the  K\"ahler potentials of $\Sigma^{[p]}$   coincide with
$-\Phi_p$,  up to an  irrelevant
re-definition  -- that do not change the Riemannian metric.
 This terminates the proof. \qed

\subsection{Toda fields from $D_n$--W-surfaces}
In this subsection we establish the following converse to
theorem \ref{TodaPlucker} and corollary
\ref{KahlerToda}
\begin{theorem}{Toda solution from Pl\"ucker embeddings.}

The K\"ahler potentials  of  any  $D_n$--W-surface introduced
by definition \ref{W-surfaces} and of its associated surfaces
introduced by definition
\ref{Associated surfaces}, may be written as
\bea
{\cal K}^{[p]}( f, \cdots ,  f^{(p-1)},
\fb, \cdots , \fb^{(p-1) })& =&
 -\Phi_{p}(z,\zb ),\quad p\leq n-1,  \nnn
{\cal K}^{[n-1]}( f, \cdots ,  f^{(n-1)},
\fb, \cdots , \fb^{(n-1) })& =&
 -(\Phi_{n-1}(z,\zb) +\xi_n +\xib_n), \nnn
{\cal K}^{[n]}( f, \cdots ,  f^{(n-2)},
\fb, \cdots , \fb^{(n-2) }) &=&
 -(\Phi_{n}(z,\zb) +\xi_n +\xib_n),
\label{kah}
\eea
where $\Phi_p$ are solutions  of the $D_n$-Toda equations.
\end{theorem}

\proof
At this point it is useful to recall the expression of the
Cartan matrix, which is   the same as for $A_n$, except  in the
following lower right $3 \times 3$ corner
\beq
\left (
\begin{array}{ccc}
 2 & -1 & -1  \\
-1 &  2 &  0  \\
-1 &  0 &  2
\end{array} \right ).
\nnn
\eeq
First  we re-derive the
Toda equation for $p\leq n-3$ directly from the fermionic expressions
obtained by substituting (\ref{wsurf}) in the K\"ahler potential
(\ref{15.31}), that is,
\beq
e^{-\p_p}=<0| \fb \cdots \fb^{(p-1)} f^{(p-1)} \cdots f |0>.
\label{15.70}
\eeq
 By explicit computations,
 one finds
\[
e^{-2\p_p}\partial \partialb \p_p=
\]\[
 <0| \fb \cdots \fb^{(p-2)}\fb^{(p)}
f^{(p-1)} f^{(p-2)} \cdots f |0>
<0| \fb \cdots \fb^{(p-2)}\fb^{(p-1)}
f^{(p)} f^{(p-2)} \cdots f |0>
\]\[
-<0| \fb \cdots \fb^{(p-2)}\fb^{(p)}
f^{(p)} f^{(p-2)} \cdots f |0>
<0| \fb \cdots \fb^{(p-1)}
f^{(p-1)} \cdots f |0>,
\]
and, thereof, applying Wick's theorem,
\[
\partial \partialb \p_p=-
e^{2\p_p} <0| \fb \cdots \fb^{(p)}
f^{(p)}  \cdots f |0>
<0| \fb \cdots \fb^{(p-2)}
f^{(p-2)} \cdots f |0>.
\]
According to the form of the Cartan matrix for $D_n$, this coincides with
Toda equations for $p\leq n-3$.
Consider, now the
case $p=n-2$. Clearly the derivation just recalled
works in the same way, but now gives   fermionic expressions
that are generalizations of expression (\ref{15.70}).
Thus we introduce
\bea
\Delta_{n-1}&\equiv & <0| \fb \cdots \fb^{(n-2)}
f^{(n-2)}  \cdots f |0>,\nnn
\Delta_n&\equiv &<0| \fb \cdots \fb^{(n-1)}
f^{(n-1)}  \cdots f |0>.
\label{15.71}
\eea
Now we show how
 they are related with  the  additional bosonic coset
spaces discussed in the subsection \ref{additional coset}.
Since $\Cco {n-1} $, and
$\Cco {\lambda_{n-1}+\lambda_n} $ (resp. $\Cco {n} $, and
$\Cco {2\lambda_n} $) have the same dimension, formulas (\ref{ass})
taken for $p=n-1$, and $p=n$, also define associated surfaces in
$\Cco {\lambda_{n-1}+\lambda_n} $, and $\Cco {2\lambda_n} $.
First, extending the preceding derivation, one
immediately sees  that
\beq
\Delta_{n-1}= e^{-(\xi_{n-1}+\xi_{n}+\xib_{n-1}
+\xib_n)} < \lambda_{n-1}+\lambda_n | \Mb^{-1} M
|\lambda_{n-1}+\lambda_n> \equiv e^{-\p_{\lambda_n+\lambda_{n-1}}}.
\label{n+n-1}
\eeq
On the other hand, and making use of (\ref{17.11}), one concludes that
\[
\Delta_{n}= e^{-2(\xi_{n-1}
+\xib_{n-1})} < 2\lambda_{n-1}| \Mb^{-1} M
|2\lambda_{n-1}>+  e^{-2(\xi_{n}
+\xib_n)}< 2\lambda_{n}| \Mb^{-1} M
|2\lambda_{n}>,
\]
\beq
\Delta_n=
e^{-\p_{2\lambda_{n-1}}}+ e^{-\p_{2\lambda_{n}}}.
\label{15.71x}
\eeq
Combining the $D_n$-Toda equations with the relations
satisfied by $\Delta_{n-1}$ and $\Delta_n$ (thanks to
Wick's theorem),  we  find  that we should have
\[
\partial\partialb \p_{n-2}=e^{2\p_{n-2}-\p_{n-3}-\p_{n-1}-\p_{n}}
=e^{2\p_{n-2}-\p_{n-3}-\p_{\lambda_n+\lambda_{n-1}}},
\]
so that
\beq
\p_{\lambda_n+\lambda_{n-1}}=\p_{n-1}+\p_{n}.
\label{15.72}
\eeq
Moreover,
\bea
\partial\partialb \p_{\lambda_n+\lambda_{n-1} }=
 \partial\partialb( \p_{n-1}+\p_n)
&=&e^{-\p_{n-2}} ( e^{2\p_{n-1}}+ e^{2\p_{n}})
\nnn
&=& e^{2\p_{\lambda_n+\lambda_{n-1}}-\p_{n-2}-\p_{2\lambda_n}},
\nonumber
\eea
so that
\beq
e^{-\p_{2\lambda_n}}=e^{-2\p_{n-1}}+e^{-2\p_{n}}.
\label{15.73}
\eeq
Expressions (\ref{15.72}) and (\ref{15.73}) are immediate consequences of
(\ref{n+n-1}) and (\ref{15.71x}), in view  of the relationship
(\ref{15.c5}) between K\"ahler potentials. \qed

As a preparation for the coming subsection, let us note that, due to the
connection between K\"ahler potentials and Toda fields
just established, it follows from the Toda equations
that the intrinsic metric tensor
$g^{[p]}_{z\zb}$ of $\Sigma^{[p]}$ is given by
\beq
g^{[p]}_{z\zb}\equiv-\partial\partialb \Phi_p=
\exp\left (\sum_j K^{(D_n)}_{pj} \>\p_j\right ).
\label{15.66x}
\eeq

\subsection{Infinitesimal Pl\"ucker formulae}
Extending the discussion of \cite{GM}, we
next show that the connection with Toda dynamics
immediately leads to the
\begin{theorem}{\bf Infinitesimal Pl\"ucker formulae.}
\label{IPlucker}

At the regular points of the embedding, the family of
scalar curvatures are related by
\begin{equation}
\label{15.68}
R_{z\, \zb}^{(p)} \sqrt{g_{z\, \zb}^{(p)}}
=\sum_q K^{(D_n)}_{pq} g_{z\, \zb}^{(q)}.
\end{equation}
\end{theorem}

\proof
This is derived by
computing  the curvature
\begin{equation}
\label{15.69}
R_{z\, \zb}^{(k)}\sqrt{g_{z\, \zb}^{(k)}}
\equiv -\partial \partialb \ln  g_{z\, \zb}^{(k)}
=-\partial \partialb \ln \left (
\exp\left (\sum_j K^{(D_n)}_{pj} \>\p_j\right )
\right ).
\end{equation}
\qed
\section{General formulation}
The consideration of the $W$-geometry of the Toda systems associated with the
algebras $C_n$ and $B_n$ follows exactly the same direction as the $D_n$-case,
and is even simpler. Before discussing the main steps of the construction
for an arbitrary simple classical Lie algebra ${\cal G}$,
let us recall briefly
some information about these two series, see e.g. \cite{B}, and their
fermionic realizations.

For the algebra $C_n$ the roots are of the form $\vec \alpha =\pm 2\vec e_p,\,
1\leq p\leq n;\; \pm \vec e_p\pm \vec e_q,\, p<q$; and the
elements of $C_n$ can be realized using $2n$
fermionic operators $\flat_{\pm p}$. The simple (positive) roots are
$\vec \pi _i=\vec e_i-\vec e_{i+1},\, 1\leq i\leq n-1$, and $\vec \pi _n=
2\vec e_n$; the corresponding fundamental weights are $\vec \lambda _i=
\sum _{1\leq j\leq i}\vec e_j,\, 1\leq i\leq n$; all $n$ fundamental
representations are of the same nature, and have the dimensions ${{2n}
\choose i }-{{2n} \choose {i-2} }$. Their weight vectors have integer
components, and are realized in the Fock space with the highest weight states
$|\lambda _i>\equiv \flat_i^+\cdots \flat_1^+|0>,\; 1\leq i\leq n$, satisfying
(\ref{A14}) with the cyclic vacuum vector $|0>$, $\flat_p|0>=
0$ for all $p=1,\cdots ,n$.

For the algebra $B_n$ the roots are $\vec \alpha =\pm \vec e_p,\,
1\leq p\leq n;\; \pm \vec e_p\pm \vec e_q,\, 1\leq p< q\leq n$; and the
root vectors corresponding to these roots are
realized in terms of $2n+1$ fermionic operators $\flat_{\pm p},\, 1\leq p\leq
n$, and $\flat_0$; the simple roots are
$\vec \pi _i=\vec e_i-\vec e_{i+1},\, 1\leq i\leq n-1$, and $\vec \pi _n=
\vec e_n$; the corresponding fundamental weights are $\vec \lambda _i=
\sum _{1\leq j\leq i}\vec e_j,\, 1\leq i\leq n-1$, and $\vec \lambda _n=
\frac{1}{2}\sum _{1\leq j\leq n}\vec e_j$. Here only the first $n-1$
fundamental representations have the weight vectors with the integer
components, and the highest weight states $|\lambda _i>\equiv \flat_i^+\cdots
\flat_1^+|0>,\; 1\leq i\leq n-1$, have the dimensions ${{2n+1} \choose i }$,
while the last one is spinorial; its dimension is $2^n$. All the reasonings
given in the previous section for the $D_n$-case work precisely in the same
way, with the relevant minor modification; here, besides $|\lambda _i>,\,
1\leq i\leq n-1$, there are two other highest states $\flat_n^+\cdots
\flat_1^+|0>$ and $\flat_{-1}^+\flat_{n-1}^+ \cdots \flat_1^+|0>$.

With these words and some
algebra, one arrives at the analogous conclusions as for the $D_n$-case,
concerning the relation between the K\"ahler potentials of the corresponding
${\cal C}^{[i]}$-manifolds and the Toda fields satisfying Eqs.(\ref{A2}) with
$K$ being the Cartan matrix of the algebra $C_n$ or $B_n$, etc. Let us only
mention that the reconstruction formulas (\ref{A9}) and (\ref{A10}) take place
for all $1\leq j\leq n$ for $C_n$ series; for $B_n$ it is valid for $j\leq
n-1$, while $e^{-2\Phi _n}=\Delta _n$.

It is natural to decouple the construction in two steps. First, let us
parametrize the cosets of $G$ for all representations of ${\cal G}=\mbox{ Lie
 }G$ with the highest weights $\lambda _p,\,p=1,\cdots ,n$, by
exponentiating the linear span of the quotient ${\cal G}/
{\cal G}^{[p]}_{\|} $. As we have already said, for all simple non-exceptional
Lie
algebras we use the fermionic realization of their elements, and  the number
of the creation $\flat^+_A$
(annihilation  $\flat_A$) operators is equal to the dimension of the Euclidean
space whose coordinates parametrize  the positive and negative roots of ${\cal
G}$. Namely, for nonexceptional representations of ${\cal G}$ we have
\begin{eqnarray}
{\cal C}^{[p]} & : & e^{\Omega _p}|\lambda _p>= e^{\sum _a{\cal
F}_a^{[p]}x_a^{[p]}}
|\lambda _p>\equiv e^{\sum _a{\cal F}_a^{[p]}x_a^{[p]}}\flat^+_p\cdots
\flat^+_1|0>
\label{A34} \\
& \equiv & \sum _{A_1,\cdots , A_p}X^{[p]p,A_p}X^{[p]p-1,A_{p-1}}
\cdots X^{[p]1,A_1}\flat^+_{A_p}\cdots \flat^+_{A_1}|0>. \label{A35}
\end{eqnarray}
Here $|0>$ is the vacuum cyclic vector, by action on which of the creation
operators $\flat^+_p\cdots \flat^+_1$ one obtains the highest weight
$\lambda _p$ state $|\lambda _p>$; $\Omega _p$ is expanded over the
elements ${\cal F}_a^{[p]}$ of ${\cal G}^{[p]}_{\bot}$;
the series in (\ref{A35})
gives a decomposition over all vectors $\flat^+_{A_p}\cdots \flat^+_{A_1}|0>
\equiv |A_p\cdots A_1>$ of the $p$-th fundamental representation
space. For the case
of the exceptional fundamental representations (the last one for $B_n$, and
the last two ones for $D_n$), the meaning of the vacuum vector $|0>$ and a
fermionic realization of the elements of the algebra is different than those
for nonexceptional representations; and formula (\ref{A35}) is modified, see
above.

The space ${\cal G}^{[p]}_{\bot}$, as an algebraic manifold,
is parametrized by independent coordinates $x_a^{[p]},\,1\leq a \leq \mbox{
dim }{\cal G}^{[p]}_{\bot}\equiv N_p$,
in the space $\tilde {\cal G}^{[p]}_{\bot}$, dual
to the space ${\cal G}^{[p]}_{\bot}$, with
the following elements: the Cartan generator $h_p$; and the root vectors
corresponding to the root string $\alpha _1^{[p]},\cdots \alpha _{N_p}^{[p]}$
containing the simple root $\pi _p\equiv \alpha _1^{[p]}$. At the same time,
the coset ${\cal C}^{[p]}=G/G_{\|}^{[p]}$, as a group manifold, is
parametrized by the coordinates $X^{[p]\alpha ,A}$ in the space
 dual to the space $G/G_{\|}^{[p]}$ corresponding, in addition to
those of ${\cal G}^{[p]}_{\bot}$, the double highest weight $\omega _p$
of the p-th fundamental representation, and all the differences
$2\omega _p - \alpha _i^{[p]}$ which do not coincide with the roots from the
root string defined above. (Of course, for the series $A_n$ the set of the
elements of ${\cal G}^{[p]}_{\bot}$
and those in the r.h.s. of (\ref{A34}) are in
one--to--one correspondence.) By this reason,
already on this step, one comes to the homogeneous
quadratic relations for the coordinates $X^{[p]\alpha ,A}$, so to
deal only with the independent  coordinates of the cosets.
Thus we arrive at a realization of the cosets in terms of the
coordinates which satisfy the relations corresponding to some algebraic
curves and surfaces. However, on the different cosets (for different values
of $p$), the coordinates $X^{[p]\alpha ,A}$
and $\bar{X}^{[p]\alpha ,A}$ clearly
are different, and are not connected yet.  And, of course, they satisfy their
own quadratic relations also separately; the origin of the relations has been
explained above.
Finally, define the K\"ahler potential ${\cal K}^{[p]}
(X^{[p]},\bar {X}^{[p]})$ of a  ${\cal C}^{[p]}$ in accordance with
(\ref{A35})as appropriate scalar product in the space of the $p$-th fundamental
representation, and recall again that, up to now, the potentials for different
manifolds ${\cal C}^{[p]}$ are defined independently.

The given reasonings clarify the origin of the quadratic relations from the
purely Lie algebraic point of view. At the same time, in the differential
geometry language, the necessity of these relations for the case of an
arbitrary simple Lie algebra ${\cal G}$ is still the decomposability of a
matrix
representative of the modified  Pl\"ucker image of ${\cal C}^{[p]}$ (for the
corresponding algebra ${\cal G}$) in the relevant subspace of the projective
space. Here, of course, one takes into account the specific of the structure
of the representation space vectors for this or that simple Lie algebra.
However, it seems to us  clearer  to formulate the relations in question
not for the right coordinates $\Lambda ^{[p]}_{A_1,\cdots , A_p}$ of the
manifold, but directly in terms of the $X^{[p]\alpha ,A}$'s, as it has been
done in the previous section for the $D_n$-case; the same is for the series
$B_n$ and $C_n$.

In fact, the homogeneous space ${\cal C}^{[p]}$ is a flag manifold
(or a parabolic space); and since we deal with $G$ being a connected complex
algebraic group, the algebraic manifold ${\cal C}^{[p]}$ naturally is a
projective and simply connected manifold. The set of the flag manifolds
${\cal C}^{[p]}$ which we consider here, realizes the cosets associated  with
the fundamental representations of ${\cal G}$, and is defined
by the corresponding parabolic subalgebras of ${\cal G}$, or, up to a local
isomorphism, by its ${\bf Z}$--gradations. The
relevant reconstruction procedure looks as follows, see e.g. \cite{VGO}.
Up to a transformation from the
inner automorphisms group $\mbox{ Int }{\cal G}=\mbox{ Ad }(G)$,
a ${\bf Z}$--gradation of ${\cal G}$ can be given by the element
$H$ from the Cartan subalgebra ${\cal H}$ of ${\cal G}$, namely,
${\cal G}_m=\{{\cal F}\in {\cal G}: [H, {\cal F}]=m{\cal F}\}$,
such that $\pi _i(h)\equiv m_i$ are nonnegative integers for all $1\leq
i\leq n$. It is clear that
\[
{\cal G}_m|_{ m\in {\bf Z}}=\delta _{m0}{\cal H}\oplus
\bigoplus_{\alpha \in \Delta_m}{\cal G}_{\alpha},\]
with
\[
\Delta_m\equiv \Delta_m(\pi_{i_1},\cdots ,\pi_{i_s})=\{\alpha =
\sum _{1\leq i\leq n}q_i\pi _i \in \Delta :
\sum _{1\leq i\leq n}q_i m_i =m\};\]
and, moreover, these subspaces ${\cal G}_{\alpha}, \alpha \in
\Delta_m$, are invariant with respect to ${\cal G}_0$. Here by
$\pi_{i_1},\cdots ,\pi_{i_s}$ we denote such the simple roots
which correspond to nonzero values of $m_i$. In accordance with this
gradation of ${\cal G}$, ${\cal G}=\oplus _{-n\leq m\leq n}{\cal G}_m$,
the subalgebra ${\cal P}^+=\oplus _{0\leq m\leq n}{\cal G}_m$,
and the opposite to it (under the reflection $\alpha \rightarrow -\alpha$)
${\cal P}^-=\oplus _{0\leq m\leq n}{\cal G}_{-m}$, are the Lie algebras
of parabolic subgroups $P^{\pm}$ of the Lie group $G$.
For the case $s=1$ these subgroups are the maximal
nonsemisimple subgroups of $G$, and just this case corresponds to
the flag manifolds ${\cal C}^{[p]}, \, p=i_1$, which realize the cosets
we are looking for.

On the second step, let us now identify  the K\"ahler potentials
${\cal K}^{[p]}(X^{[p]},\bar{X}^{[p]})$  with the Toda fields satisfying
the equations of motion, just by setting
\[
e^{\Omega _p}|\lambda _p>= Me^{-\sum_{ij}h_i(k^{-1})_{ij}\mbox{
log }s_j}|\lambda _p>;\]
cf.(\ref{A3}), on the corresponding $W_{\cal G}$ - and associated surfaces.
Here arises the first nontrivial point. With this identification, the
coordinates $x_a^{[p]}$ for different values of $p$ are not already
independent,and are constructed in terms of the same screening functions
$s _j$. So, one should get convinced in the following two statements:

i) The functions $f^A= X^{[1]1,A}$ entering ${\cal C}^{[1]}$ do satisfy
the corresponding quadratic relations; in other words these relations do not
contradict to the nested structure (\ref{15.50}) of  $f^A$.

ii) The functions $f^{(p-1)A}=X^{[p]\alpha ,A}$, $p>1$, entering
${\cal C}^{[p]}$, lead to the Toda fields $\Phi _p$ determined by formulas
(\ref{A3}) via the screening functions, and satisfy the same quadratic
relations as above.

In the previous section we have proved these statements for the $D_n$-case by
a direct verification; for $B_n$ and $C_n$ series it can be shown in the
similar way. So, the relations in question are identically satisfied on the
class of the solutions to the Toda system, when the coordinates
$X^{[p]\alpha ,A}$ are expressed via the screening functions as the nested
integrals (\ref{15.50}), i.e. on the corresponding $W$-surfaces.

Of course, our discussion of the quadratic relations concerns only a part
of the problem. We have the cosets ${\cal C}^{[p]}$
which, in general, are submanifolds of the projective spaces. The holomorphic
(antiholomorphic) blocks entering the K\"ahler potentials under their
identification with the Toda fields $\Phi _i(z,\bar{z})$ given by formula
(\ref{A3}), are related to the nested integral structure of the nilpotent
elements $M$ ($\bar{M}$) written in terms of the screening functions,
\begin{equation}
M=\sum_{m=1}^{\infty}\sum_{1\leq i_1,\cdots ,i_m\leq n}
(i_1,\cdots ,i_m)E_{-i_m}\cdots E_{-i_1}, \label{A36}
\end{equation}
where $(i_1,\cdots ,i_m)$ is the compact notation (\ref{15.50}) for the
corresponding repeated integrals.

Just this representation automatically takes into account the fact that the
functions $f^A$ are not independent, and are parametrized by exactly $n$
number of independent screening functions $s_i$. The embedding
functions $f^A(z)$ ($\bar{f}^A(\bar{z})$) which define the corresponding
$W_{\cal G}$ - and associated surfaces satisfy the necessary quadratic
relations just thanks to their nested structure. One can move in the opposite
direction and observes that the identification
of the Toda fields with the K\"ahler potentials for the associated surfaces
in ${\cal C}^{[p]}$ gives, that the embedding functions $f^A(z)$
($\bar{f}^A(\bar{z})$) and their derivatives of the corresponding order,
coincide with the coordinates $X^{[p]\alpha ,A}$, cf. with (\ref{A27}), and
provides the necessary relations. And, moreover, the K\"ahler potentials of
the manifolds ${\cal C}^{[p]}$ satisfy the system of partial differential
equations (\ref{A2}).

Finally, the second part of the
formula (\ref{A31}) for the case of a simple Lie algebra ${\cal G}$ endowed
with the principal gradation, takes, with account of the equations of motion
(\ref{A2}), the form
\begin{equation}
d{\cal S}^2_k\,=\,\frac{i}{2}\exp 2(\sum _{j=1}^nK_{kj}\Phi _j)\,
dz\, d\bar {z}\,\equiv \,\frac{i}{2}\exp 2\rho _j\,
dz\, d\bar {z}.
\label{A38}
\end{equation}
The curvature form of the pseudo-metrics $d{\cal S}^2$ appears as
\begin{equation}
-i \partial \bar {\partial }\rho _k \,=\,\sum
_jK_{kj}d{\cal S}^2_j.
\label{A39}
\end{equation}
Then we naturally come to the following  concerning a
generalization for an arbitrary simple Lie algebra ${\cal G}$
of the global Pl\"ucker formula
\begin{conjecture}{Global Pl\"ucker formula.}

For an arbitrary simple Lie algebra ${\cal G}$, with
 degrees $d_k$, on a W-surface of   genus $g$,
with  total ramification numbers  $\beta _k$,
 one has
\begin{equation}
2g-2-\beta _k+2\sum _{j=1}^nK_{kj}d_j=0.
\label{A40}
\end{equation}
\end{conjecture}

In accordance with an interpretation given in \cite{GM}, $W$-surfaces for the
case of $A_n$ are instantons of the associated nonlinear $\sigma $-model, and
in turn are described by the solutions of the cylindrically symmetric
self--dual Yang-Mills equations, for which the action coincides, up to
inessential
numerical factor, with the topological charge (or Pontryagin index, or
instanton number) $Q_k$ of this configuration. The same reasonings work
also for the cylindrically symmetric self--dual fields associated with an
arbitrary simple Lie algebra ${\cal G}$ which, in accordance with \cite{LS},
satisfy the Toda system of equations (\ref{A2}). Here there is also an
explicitexpression for the topological charge density, which provides, with the
help
of the Gauss--Bonnet formula, a bridge between the infinitesimal (\ref{A39})
and  the global (\ref{A40}) Pl\"ucker type formulas. In other words, formula
(\ref{A40}) gives a relation between the genus of a $W_{\cal G}$ - manifold
and its topological characteristics $Q_k=d_k$. Moreover, since the
cylindrically symmetric instantons for ${\cal G}$ constitute a subclass of
$2r$-parametric solutions of (\ref{A2}) regular on the one-point
compactification of ${\bf R}^4$ and with finite action (or topological
charge),a justification of these requirements by imposing the corresponding
boundary
conditions on the Toda fields, leads to the evident relation between the
ramification indices $\beta _k$ and the degrees $m_k$ of the singularities of
the functions $\exp 2\rho _j$ in
the r.h.s of (\ref{A38}), $\beta _i =\sum _{j=1}^n\,K_{ij}\,m_j$.
With such a standpoint, the integers $m_k$ are nothing but the integration
constants entering the parametrization $s_i(z)=c_{i}\exp (m_iz)$,
$\bar{s}_i(\bar{z})=\bar{c}_{i}\exp (m_i\bar{z})$ of the arbitrary
screening functions
$s_i (z)$ and $\bar{s}_i (\bar{z}),\,1\leq i\leq n$, which determine the
general solutions (\ref{A3}) of the Toda system (\ref{A2}).

\section{Acknowledgements}
We are indebted to Y. Matsuo for his collaboration at an early stage of this
work. We would like to thank D. V. Alexeevskii, B. L. Feigin, P.
Gauduchon, B. A. Khesin, M. L. Kontsevich, Yu. I. Manin, and A. V. Razumov
for the very useful discussions; and F. E. Burstall,
Ph. A. Griffiths, and L. M. Woodward for interesting communications.
It is a great pleasure
for one of the
authors (M.V.S.) to acknowledge the warm hospitality and
 creative
scientific atmosphere of the Laboratoire de Physique Th\'eorique de
l'\'Ecole Normale Sup\'erieure de Paris.
This work was supported in part by the EEC grant \# SC1*-0394-C.

\appendix

\section{Appendix: Group properties for $D_n$}
\subsection{The bosonic representations}
Let us determine $\Gcc p $ and $\Gcp p $ at once
 for $1\leq p \leq n-2$, where, according to
Eq.\ref{15.6}, $|\lambda_p>=\flat^+_p \flat^+_{p-1}\cdots \flat^+_1|0>$.
Let us  call ${\cal N}_+$ the niloptent Lie
algebra generated by the step
operators with positive roots.
It is easy to see that  $\Gcc p $ is given by
$$
\Gcc p = {\cal N}_+;  \left\{ h_i, i\not= p\right \};
\left\{ E_{-\vec e_\alpha+\vec e_\beta}; \alpha, \beta \leq p \right \};
$$
\beq
\left\{ E_{-\vec e_k- \vec e_\ell}; \ell >k> p \right \};
\left\{ E_{-\vec e_k+ \vec e_\ell}; \ell >k> p \right \}.
\label{15.8}
\eeq
These generators may be reorganized as follows
$$
\Gcc p = \left\{ E_{\vec e_\alpha+\vec e_\beta};
\alpha, \beta \leq p \right \};
\left\{ E_{\vec e_\alpha+ \vec e_k}; \alpha\leq p,  k> p \right \};
\left\{ E_{\vec e_\alpha- \vec e_k}; \alpha\leq p,  k> p \right \};
$$
$$\left\{ h_\alpha; \alpha\leq  p-1\right \};
\left\{ E_{-\vec e_\alpha+\vec e_\beta}; \alpha< \beta \leq p \right \};
\left\{ E_{\vec e_\alpha-\vec e_\beta}; \alpha< \beta \leq p \right \};
$$
$$\left\{ h_k; k>   p\right \};
\left\{ E_{-\vec e_k- \vec e_\ell}; \ell >k> p \right \};
\left\{ E_{-\vec e_k+ \vec e_\ell}; \ell >k> p \right \}.
$$
\beq
\left\{ E_{\vec e_k- \vec e_\ell}; \ell >k> p \right \};
\left\{ E_{\vec e_k+ \vec e_\ell}; \ell >k> p \right \}.
\label{15.9}
\eeq
The first   line  generates  a nilpotent algebra  denoted
${\cal N}_+^{[p] }$ of dimension $2p(n-p)+p(p-1)/2=2np -p(3p+1)/2$.
The next  line  clearly generates  $A_{p-1}$, which has
dimension $(p^2-1)$.
The remaining lines generate $D_{n-p}$, of dimension
$(n-p)(2n-2p-1)$. The dimension of $\Gcc p $ is
therefore $2np -p(3p+1)/2+p^2-1+(n-p)(2n-2p-1)$,
that is,   $2n^2-n-2np+3p^2/2+p/2-1$.

Next,  this  coset is parametrized by exponentiating
$$\Gcp p = \left \{h_p \right \};  \left \{ E_{-\vec e_\alpha+ \vec e_k},
, \alpha\leq p,\, k>p \right \};
 \left \{ E_{-\vec e_\alpha- \vec e_k},
, \alpha\leq p,\, k>p  \right \};
$$
\begin{equation}
\left \{ E_{-\vec e_\alpha- \vec e_\beta},
, \alpha\leq \beta \leq p  \right \}.
\label{15.11}
\end{equation}

The dimension is $2np-p(3p+1)/2+1$.  \footnote{ for $p=1$ this is actually
$2n-1$ which coincides with the dimension of
$D_n/B_{n-1}$.} Adding the dimensions of
$\Gcc p $ and $\Gcp p $ correctly gives $n(2n-1)$ which is the
dimension of $D_n$.

\subsection{Explicit parametrization of $\Cco p $ for bosonic
representations}

We treat the generic case $1\leq p\leq n-2$, where
 $|\lambda_p>=\flat^+_p\cdots  \flat^+_1 |0>$.
 According to Eq.\ref{15.11}, $\Cco p $
is parametrized by\footnote{Here again we leave aside the bar
components which are similar.}
$e^{\kappa_p  h_p} e^{\Omega_p} \flat_p^+\cdots \flat_1^+ | 0 >$,
with
\beq
\Omega_p=\sum_{\ell >p,}\sum_{\gamma \leq p}
\left (x_{\ell}^{[p]  \gamma} E_{-\vec e_\gamma+ \vec e_\ell}
+x_{-\ell}^{[p] \gamma} E_{-\vec e_\gamma-\vec e_\ell}\right )
+\sum_{1\leq \gamma<\delta\leq p}
u^{[p]}_{\gamma,\delta} E_{-\vec e_\gamma-\vec e_\delta}.
\label{15.25}
\eeq
The coset parameters are $x_{\ell}^{[p]\gamma}$,
$x_{-\ell}^{[p] \gamma}$
and $u^{[p]}_{\gamma,\delta}$.
Turning the same crank as for the first fundamental
representation, one computes (for $\alpha< p$)
$$
e^{\Omega_p} \flat_\alpha^+ e^{-\Omega_p} =
\flat_\alpha^++  \sum_{k> p,\, \epsilon=\pm 1 }
x_{\epsilon k}^{[p]\alpha} \flat^+_{\epsilon k} +
$$

\beq
\left ( \sum_{p\geq \beta>\alpha} u^{[p]}_{\alpha, \beta}
-\sum_{p\geq \alpha > \beta} u^{[p]}_{\beta, \, \alpha }\right )
\flat^+_{-\beta}
-{1\over 2} \sum_{k> p,\, \epsilon=\pm 1 }
\sum_{\gamma \leq p}
x_{\epsilon k}^{[p] \alpha} x_{-\epsilon k}^{[p]\gamma} \flat^+_{-\gamma}.
\label{15.26}
\eeq
Thus we obtain
\beq
e^{\kappa_p h_p} e^{\Omega_p} |\lambda_p>
=\sum_{A_1,\, \cdots ,  \,A_p}
X^{[p] \,p,\, A_p} X^{[p]\,p-1, \, A_{p-1}} \cdots   X^{[p]\, 1,\, A_1}
\flat_{A_p}^+ \cdots \flat_{A_1}^+ |0>,
\label{15.27}
\eeq
where for $1\leq \alpha \leq p$, $1\leq \beta \leq p$,
and $k >p$,
$$
X^{[p] \, \alpha,\, \beta } =\delta_{\alpha,\, \beta},
\quad \hbox{for}\> \alpha, \beta <p;
$$
$$ X^{[p] \, p, \, \beta }=
X^{[p] \, \beta, \, p }= e^{\kappa_p} \delta_{p,\, \beta};
$$
$$
X^{[p] \, \alpha,\, -\beta }=\left \{  \begin{array}{cc}
-{1\over 2} \sum_{k>p,\,\epsilon=\pm}\> (x_{\epsilon k}^{[p]  \alpha}
x_{\epsilon k}^{[p]  \beta})+ u^{[p]}_{\alpha,\, \beta}, &
\hbox{if}\> p >\beta > \alpha \\
-e^{-\kappa_p}{1\over 2} \sum_{k>p,\,\epsilon=\pm}\>
(x_{\epsilon k}^{[p]  \alpha}
x_{\epsilon k}^{[p]  p})+ e^{-\kappa_p} u^{[p]}_{\alpha,\, p},  &
\hbox{if}\> \beta=p, p> \alpha   \\
-{1\over 2} \sum_{k>p,\,\epsilon=\pm}\>( x_{\epsilon k}^{[p]  \alpha}
x_{\epsilon k}^{[p]  \beta})- u^{[p]}_{\beta, \, \alpha}, &
\hbox{if}\> \alpha >\beta\\
-{1\over 2} \sum_{k>p,\,\epsilon=\pm}\> (x_{\epsilon k}^{[p]  \alpha}
x_{\epsilon k}^{[p]  \beta}),  &
\hbox{if}\> \alpha =\beta \not=p\\
-e^{-\kappa_p}{1\over 2} \sum_{k>p,\,\epsilon=\pm}\> (x_{\epsilon k}^{[p]  p}

x_{\epsilon k}^{[p]  p}),  &
\hbox{if}\> \alpha =\beta=p
\end{array}\right.
$$
\beq
X^{[p] \, \alpha,\, \pm (p+1) } = e^{\mp \kappa_p}
x_{\pm (p+1)}^{[p]\alpha},
\quad  X^{[p] \, \alpha,\, \pm k} = x_{\pm k}^{[p]\alpha}, \, k>p+1.
\label{15.28}
\eeq
It is easy to see that the coordinates $X^{[p] \, \alpha,\, A}$
satisfy the polynomial conditions,
\beq
\sum_{A} X^{[p] \, \alpha,\, A} X^{[p]\, \beta,\, -A}=0,
\quad 1\leq \alpha \leq p,\>
\quad 1\leq \beta  \leq p.
\label{15.29x}
\eeq
 This explicit computations gives a parametrization
 such that
for $1\leq A\leq p$, and $1\leq \Ab\leq p$,
\beq
X^{[p]  \alpha, \, A}=\delta_{\alpha,\, A}.
\label{15.33}
\eeq
With this parametrization, we may easily
solve the constraints Eq.\ref{15.29},  and write
(for $0\leq \alpha, \beta \leq p$),
\beq
X^{[p]  \alpha, \, -\beta} +X^{[p]  \beta, \, -\alpha}=-
\sum_{A=p+1}^n X^{[p] \, \alpha,\, A} X^{[p]\, \beta,\, -A},
\label{15.34}
\eeq
so that the independent components are $X^{[p] \, \alpha,\, \pm A}$,
$\Xb^{[p] \, \alpha,\, \Ab}$, $A, \Ab \geq p+1$, and
$X^{[p]  \alpha, \, -\beta} -X^{[p]  \beta, \, -\alpha}$,
$\Xb^{[p]  \alpha, \, -\beta}- \Xb^{[p]  \beta, \, -\alpha}$.

\subsection{The three additional bosonic representations}
The previous description of
the
cosets
extends to the representations generated by $|\lambda_{n-1}+\lambda_n>$,
$|2\lambda_n>$, and $|2\lambda_{n-1}>$
without any problem. The first two cases are direct extensions of the
formulae given for $\lambda_p$ with $p\leq n-1$.  The last one is obtained
from the calculation for $2 \lambda_{n-1}$, by exchanging everywhere
$\flat_n$ with $\flat_{-n}$.
Using  similar
notations, we  introduce
\bea
\Omega_{\lambda_{n-1}+\lambda_n}&=&\sum_{\gamma \leq  n-1,}
\Bigl  (x_{n}^{[\lambda_{n-1}+\lambda_n]  \gamma} E_{-\vec e_\gamma+ \vec e_n}
\nnn
&\> &+x_{-n}^{[\lambda_{n-1}+\lambda_n] \gamma} E_{-\vec e_\gamma-\vec
e_n}\Bigr )
+\sum_{1\leq\gamma<\delta\leq n-1}
u^{[\lambda_{n-1}+\lambda_n]}_{\gamma,\delta}
E_{-\vec e_\gamma-\vec e_\delta}, \nnn
\Omega_{2\lambda_n}&=&
\sum_{1\leq \gamma<\delta\leq n} u^{[2\lambda_n]}_{\gamma,\delta}
E_{-\vec e_\gamma-\vec e_\delta} \nnn
\Omega_{2\lambda_{n-1}}&=&\sum_{\gamma \leq  n-1,}
x_{n}^{[2\lambda_{n-1}]  \gamma} E_{-\vec e_\gamma+ \vec e_n}
+\sum_{1\leq \gamma<\delta\leq n-1}
u^{[2\lambda_{n-1}]}_{\gamma,\delta}
E_{-\vec e_\gamma-\vec e_\delta}.
\nonumber
\eea
\beq
X^{[\lambda]  r} =e^{\kappa_\lambda h_\lambda} e^{\Omega_\lambda} \flat_r^+
e^{-\Omega_\lambda} e^{-\kappa_\lambda h_\lambda}
\label{15.33r}
\eeq
\subsection{The fermionic representations}
The corresponding coset manifolds are studied in the same way as
above. They are parametrized by
$e^{\kappa_p  h^{(1/2)}_p} e^{\Omega^{(1/2)}_p} |\lambda_p>$,
where
\bea
\Omega^{(1/2)}_{n-1}&=&\sum_{\gamma }
x_{n}^{[n-1]  \gamma} E^{(1/2)}_{-\vec e_\gamma+ \vec e_n}
+\sum_{1\leq \gamma<\delta\leq n-1} u^{[n-1]}_{\gamma,\delta}
E^{(1/2)}_{-\vec e_\gamma-\vec e_\delta},  \nnn
\Omega^{(1/2)}_{n}&=&
\sum_{1\leq \gamma<\delta} u^{[n]}_{\gamma,\delta}
E^{(1/2)}_{-\vec e_\gamma-\vec e_\delta}.
\label{15.g}
\eea

\end{document}